\documentstyle[aaspp4,epsfig]{article}

\newcommand{\vv}[1]{{\bf #1}}

\newcommand{\avg}[1]{{\langle{#1}\rangle}}

\def\simless{\mathbin{\lower 3pt\hbox
	{$\,\rlap{\raise 5pt\hbox{$\char'074$}}\mathchar"7218\,$}}} 
\def\simgreat{\mathbin{\lower 3pt\hbox
	{$\,\rlap{\raise 5pt\hbox{$\char'076$}}\mathchar"7218\,$}}} 

\newcounter{thefigs}
\newcommand{\fignum}{\arabic{thefigs}}

\newcounter{thetabs}
\newcommand{\tabnum}{\arabic{thetabs}}

\begin{document} 

\title{Time Evolution of Galaxy Formation \\
and Bias in Cosmological Simulations}
\vspace{20pt}
\author{Michael~Blanton, Renyue~Cen, \\
Jeremiah~P.~Ostriker, Michael~A.~Strauss}
\affil{Princeton University Observatory, Princeton, NJ 08544 }
\affil{ blanton, cen, jpo, strauss@astro.princeton.edu}
\and
\author{Max Tegmark}
\affil{Institute for Advanced Study, Princeton, NJ 08540 }
\affil{ max@sns.ias.edu}
\vspace{20pt}

\begin{abstract} 

The clustering of galaxies relative to the underlying mass distribution
declines with cosmic time for three reasons. First, nonlinear peaks become less
rare events as the density field evolves.  Second, the densest regions stop
forming new galaxies because their gas becomes too hot to cool and
collapse. Third, after galaxies form, they are subject to the same
gravitational forces as the dark matter, and thus they tend to trace the dark
matter distibution more closely with time; in this sense, they are
gravitationally ``debiased.''  In order to illustrate these effects, we perform
a large-scale hydrodynamic cosmological simulation of a $\Lambda$CDM model with
$\Omega_0=0.37$ and examine the statistics of $\delta_\ast(\vv{r},z)$, the
density field of recently formed galaxies at position $\vv{r}$ and redshift
$z$.  We find that the bias of recently formed galaxies $b_\ast \equiv
\avg{\delta_\ast^2}^{1/2}/\avg{\delta^2}^{1/2}$, where $\delta$ is the mass
overdensity, evolves from $b_\ast\sim 4.5$ at $z=3$ to $b_\ast\sim 1$ at $z=0$,
on 8 $h^{-1}$ Mpc comoving scales. The correlation coefficient $r_\ast \equiv
\avg{\delta\delta_\ast}/\avg{\delta^2}^{1/2}\avg{\delta_\ast^2}^{1/2}$ evolves
from $r_\ast\sim 0.9$ at $z=3$ to $r_\ast\sim 0.25$ at $z=0$. That is, as gas
in the universe heats up and prevents star formation, the star-forming galaxies
become poorer tracers of the mass density field.  We show that the linear
continuity equation is a good approximation for describing the gravitational
debiasing, even on nonlinear scales.  The most interesting observational
consequence of the simulations is that the linear regression of the galaxy
formation density field on the galaxy density field, $b_{\ast g}r_{\ast g} =
\avg{\delta_\ast\delta_g}/\avg{\delta_g^2}$, evolves from about 0.9 at $z=1$ to
0.35 at $z=0$.  Measuring this evolution, which should be possible using the
Sloan Digital Sky Survey, would place constraints on models for galaxy
formation.  In addition, we evaluate the effects of the evolution of galaxy
formation on estimates of $\Omega$ from cluster mass-to-light ratios, finding
that while $\Omega(z)$ increases with $z$, the estimated
$\Omega_{\mathrm{est}}(z)$ actually decreases. This effect is due to the
combination of galaxy bias and the relative fading of cluster galaxies with
respect to field galaxies. Finally, these effects provide a possible
explanation for the Butcher-Oemler effect, the excess of blue galaxies in
clusters at redshift $z\sim 0.5$.

\end{abstract}

%
%
%
\setcounter{section}{0}
\section{Motivation}
\label{motiv}

Recent observations of the evolution of galaxy clustering have
explored both the regime
$z \simless 1$, and have probed the distribution of galaxies at $z
\approx 3$.  The low-redshift works include a wide-field ($4^\circ
\times 4^\circ$) $I$-selected angular sample (\cite{postman98ap}),
Canada-France Redshift Survey galaxies with redshifts between $0 < z <
1.3$ (\cite{lefevre96ap}), the Hubble Deep Field
(\cite{connolly98ap}), field galaxies from the Canadian Network for
Observational Cosmology (CNOC) survey (\cite{shepherd97ap}), the
Hawaii $K$-selected redshift survey (\cite{carlberg97ap}), and a
redshift survey performed on the Palomar 200-inch telescope
(\cite{small99ap}).  The results of these surveys are still rather
discrepant; they don't yet give a consistent picture for how galaxy
clustering evolves with time. 
The evolution of the relative clustering of old and
and young galaxies is also of interest, because it yields direct
information on where galaxy formation occurs as a function of
redshift. \cite{lefevre96a} made a simple attempt at measuring this
evolution by separating their sample by color. They found that red and
blue galaxies have comparable correlation amplitudes at $z>0.5$,
although red galaxies are more clustered today, by a factor of about
1.5 (in agreement with other local estimates, such as
\cite{davis76ap}, \cite{giovanelli86ap}, \cite{santiago92ap},
\cite{loveday95ap}, \cite{loveday96ap}, \cite{hermit96ap}, and
\cite{guzzo97ap}).  The evolution of the density-morphology relation
in clusters also contains information about where galaxies form as a
function of redshift, showing that at $z\sim 0.5$ there is
significantly more star formation in clusters than there is today
(Butcher \& Oemler 1978, 1984; \cite{couch87ap}; \cite{dressler92ap};
\cite{couch94ap}; \cite{oemler97ap}; \cite{dressler97ap};
\cite{poggianti99ap}).

Measurements of the clustering of Lyman-break objects (LBOs) at $z\sim 3$
(\cite{steidel98ap}) give a more unambiguous indication of the
evolution in clustering.  In particular, measurements of the amplitude of
counts-in-cells fluctuations (\cite{adelberger98ap}) or the angular
autocorrelation function (\cite{giavalisco98ap}) 
suggest that galaxies were as strongly clustered in comoving
coordinates at $z \sim 3$ as they are today.  If the LBOs are unbiased
tracers of the mass density field, these results contradict the widely accepted
gravitational instability (GI) model for the formation of large-scale
structure, unless one assumes an unacceptably low value of $\Omega_0$ in order
to prevent the growth of the clustering of galaxies between $z=3$
and today.  Therefore, the objects observed at high redshift are probably more
highly ``biased'' tracers of the mass density field than are galaxies
today. That is, they have a high value of $b_g\equiv\sigma_g/\sigma$, where
$\sigma_g\equiv\avg{\delta_g^2}^{1/2}$ is the rms galaxy density fluctuation
and $\sigma\equiv\avg{\delta^2}^{1/2}$ is the rms mass fluctuation. The
counts-in-cells analysis of \cite{adelberger98a} suggests that
at $z=3$ the bias of LBOs is $b_g \sim 2 b_0$ (for $\Omega_0=0.2$,
$\Omega_\Lambda = 0$), $b_g \sim 4 b_0$ (for $\Omega_0=0.3$, $\Omega_\Lambda =
0.7$), or $b_g\sim 6 b_0$ (for $\Omega_0=1$), where $b_0$ is the bias of
galaxies today. 

How does the observed clustering since $z\approx 3$ relate to the underlying
mass density fluctuations? From a theoretical perspective, the bias decreases
with time because of three effects. First, at early times, collapsed objects
are likely to be in the highest peaks of the density field, since one needs a
dense enough clump of baryons in order to start forming stars.  Such
high-$\sigma$ peaks are highly biased tracers of the underlying density field
(\cite{doroshkevich70ap}; \cite{kaiser84ap}; \cite{bardeen86ap};
\cite{bond91ap}; \cite{mo96ap}).  As time progresses, lower peaks in the
density field begin to form galaxies, which are less biased tracers of the mass
density field. Second, the very densest regions soon become filled with
shock-heated, virialized gas which does not easily cool and collapse to form
galaxies (\cite{blanton98ap}). These two effects cause a shift of galaxy
formation to lower density regions of the universe. This shift is evident in
the real universe: young galaxies are not found in clusters at $z=0$.  To keep
track of these effects, we define the bias $b_\ast$ and correlation
coefficient $r_\ast$ of galaxy formation as
\begin{equation}
b_\ast \equiv \frac{\avg{\delta_\ast^2}^{1/2}}{\avg{\delta^2}^{1/2}} \quad 
{\mathrm{and}}\quad
r_\ast \equiv \frac{\avg{\delta_\ast\delta}}{\avg{\delta_\ast^2}^{1/2}
\avg{\delta^2}^{1/2}},
\end{equation}
where $\delta\equiv\rho/\avg{\rho}-1$ is the overdensity of mass and
$\delta_\ast\equiv\rho_\ast/\avg{\rho_\ast}-1$ is the overdensity field of
``recently formed'' galaxies, each defined at some smoothing scale. By
``recently formed,'' we mean material that has collapsed and formed stars
within the last 0.5 Gyrs. Alternatively, one can consider $\delta_\ast$ as the
overdensity field of galaxies weighted by the star formation rate in each ---
for the purposes of this paper we consider the formation of a galaxy equivalent
to the formation of stars within it. As functions of redshift, $b_\ast(z)$ and
$r_\ast(z)$ are essentially differential quantities which keep track of where
stars are forming at each epoch. They both decrease with time for the reasons
stated above. Predictions of $b_\ast(z)$ and $r_\ast(z)$ are complementary to
predictions of the star formation rate as a function of redshift
(\cite{madau96ap}; \cite{nagamine99ap}; \cite{somerville98bp};
\cite{baugh98ap}); whereas those studies examine {\it when} galaxies form, we
examine {\it where} they form.

The third effect is that after galaxies form, they are subject to the same
gravitational physics as the dark matter; thus, the two distributions become
more and more alike, debiasing the galaxies gravitationally (\cite{fry96ap};
\cite{tegmark98ap}; see Section \ref{burst}). The history of $b_\ast(z)$ and
$r_\ast(z)$ convolved with effects of gravitational debiasing determines the
history of the bias of all galaxies, quantified as
\begin{equation}
b_g \equiv \frac{\avg{\delta_g^2}^{1/2}}{\avg{\delta^2}^{1/2}} \quad 
{\mathrm{and}}\quad
r_g \equiv \frac{\avg{\delta_g\delta}}{\avg{\delta_g^2}^{1/2}
\avg{\delta^2}^{1/2}},
\end{equation}
where $\delta_g\equiv\rho_g/\avg{\rho_g}-1$ is the overdensity of
galaxies.  Here, $b_g(z)$ and $r_g(z)$ are integral quantities, which
refer to the entire population of galaxies.  
Until recently, most
papers on this subject have considered only the effect of bias ($b_g$)
on large-scale structure statistics. However, observational and
theoretical arguments suggest that scatter in the relationship
between galaxies and mass may also be important (\cite{tegmark98bp};
\cite{blanton98ap}); treating the relationship as purely deterministic
when analyzing peculiar velocity surveys or redshift-space distortions
of the power spectrum can produce observational inconsistencies
(\cite{pen98ap}; \cite{dekel98ap}).  Thus, even at lowest order, these
treatments need to include consideration of the correlation
coefficient $r_g$.

Finally, we will also define the cross-correlation of all galaxies and of young
galaxies as 
\begin{equation}
b_{\ast g} \equiv
\frac{\avg{\delta_\ast^2}^{1/2}}{\avg{\delta_g^2}^{1/2}} 
\quad {\mathrm{and}}\quad
r_{\ast g} \equiv \frac{\avg{\delta_\ast\delta_g}}
{\avg{\delta_\ast^2}^{1/2} \avg{\delta_g^2}^{1/2}}.
\end{equation}
These quantities are important because they are potentially
observable, by comparing number-weighted and star-formation weighted galaxy
density fields. 

Previous theoretical efforts have focused on the evolution of the
clustering of all galaxies, or equivalently, the evolution of
the integral quantities $b_g(z)$ and $r_g(z)$. There has been notable
success in this vein in explaining the nature of the LBOs. In the
peaks-biasing formalism, the halo mass of collapsed objects determines
both their number density and their clustering strength
(\cite{doroshkevich70ap}; \cite{kaiser84ap}; \cite{bardeen86ap};
\cite{bond91ap}; \cite{mo96ap}); interestingly, halos with masses $>
10^{12} M_\odot$ at $z=3$ have both a number density and clustering
strength similar to those of LBOs, in reasonable cosmologies
(\cite{adelberger98ap}), a result verified by $N$-body simulations
(\cite{wechsler98ap}; \cite{kravtsov98ap}). This fact is reflected in
the results for high-redshift clustering of semi-analytic galaxy
formation models, which also find that galaxies at $z\sim 3$ are
highly biased (\cite{kauffmann97ap}; \cite{kauffmann98ap};
\cite{somerville98ap}; \cite{baugh98ap}). Furthermore, hydrodynamic
simulations using simple criteria for galaxy formation
(\cite{evrard94ap}; \cite{katz98ap}; \cite{cen98ap}) find that the
distribution of galaxies at $z=3$ is indeed biased with respect to the
mass distribution, and that the clustering strength of galaxies
depends only weakly on redshift.

While the issue of $b_g(z)$ and $r_g(z)$ is the one which observations
are currently best-suited to address, perhaps other statistics can
constrain the nature of galaxy formation more powerfully. After all,
the fundamental prediction of theories for galaxy formation, such as
ours and such as the semi-analytic models mentioned above, is the
location of star formation as a function of time. Important
information about galaxy formation may be lost if one examines only
the integral quantities, which have been affected by the entire
history of galaxy formation convolved with their subsequent
gravitational evolution.
Therefore, we focus here on the evolution of the large-scale
clustering of galaxy {\it formation}, defined as the formation of the
galaxy's constituent stars --- that is, the evolution of $b_\ast$ and
$r_\ast$. Now is an opportune time to investigate such questions,
since large redshift and angular surveys which can examine this and
related properties of galaxies are imminent.

In this paper, we measure the clustering of galaxy formation in the
hydrodynamical simulations of \cite{cen98a}, which we describe in Section
\ref{simulations}. In Section \ref{timeformation}, we consider the properties
of galaxy formation in these simulations as a function of time, studying the
first two effects discussed above and the redshift dependence of $b_\ast$ and
$r_\ast$.  In Section \ref{burst}, we discuss the gravitational debiasing, the
third effect mentioned above. In Section \ref{observables}, we discuss
observable effects of the trends in galaxy formation found in the
simulations. Especially important is the strong evolution of the
cross-correlation between the star-formation weighted density field and the
galaxy density field; this evolution should be observable with surveys
such as the Sloan Digital Sky Survey (SDSS; \cite{gunn95ap}). In addition, the
evolution of the location of galaxy formation in the universe may be related to
the observed Butcher-Oemler effect and may affect the mass-to-light ratios of
clusters.  We conclude in Section \ref{conclusions}.

%
%
\setcounter{section}{1}
\section{Simulations}
\label{simulations}

The work of \cite{ostriker95a} motivated the choice of a flat cold
dark matter cosmology for the simulation used in this paper, with
$\Omega_0 = 0.37$, $\Omega_\Lambda = 0.63$, and
$\Omega_b=0.049$. Recent observations of high redshift supernovae have
lent support to the picture of a flat, low-density universe
(\cite{perlmutter97ap}; \cite{garnavich98ap}). Great uncertainty
remains, however, and future work along the lines of this paper will
need to address different cosmologies. The
Hubble constant was set to $H_0 = 100$~$h$~km~s$^{-1}$~Mpc$^{-1}$, with
$h=0.7$.  The primordial perturbations were adiabatic and random
phase, with a power spectrum slope of $n = 0.95$ and amplitudes such
that $\sigma_8=0.8$ for the dark matter at $z=0$, at which time the
age of the universe is 12.7 Gyrs. We use a periodic box 100 $h^{-1}$
Mpc on a side, with $512^3$ grid cells and $256^3$ dark matter
particles. Thus, the dark matter mass resolution is about $5\times
10^9 $ $h^{-1}$ $M_\odot$ and the grid cell size is $\sim$ 0.2
$h^{-1}$ Mpc.  
The smallest smoothing length we consider is a 1 $h^{-1}$ Mpc radius top hat,
which is considerably larger than a cell size. On these scales and larger, the
relevant gravitational and hydrodynamical physics are correctly handled.  On
the other hand, subgrid effects such as the fine grain structure of the gas and
star formation may influence large-scale properties of the galaxy
distribution. As we describe here, we handle these effects using crude, but
plausible, rules.

\cite{cen98a} describe the hydrodynamic code in detail; it is similar
to but greatly improved over that of Cen \& Ostriker (1992a,b).  
The simulations
are Eulerian on a Cartesian grid and use the Total Variation
Diminishing method with a shock-capturing scheme (\cite{jameson89ap}).
In addition, the code accounts for cooling processes and incorporates
a heuristic galaxy formation criterion, whose essence is as follows:
if a cell's density is high enough, if the cooling time of the gas in
it is shorter than its dynamical time, if it contains greater than the
Jeans mass, and if the flow around that cell is converging, it will
have stars forming inside of it. The code turns the baryonic fluid
component into collisionless stellar particles (hereafter ``galaxy
particles'') at a rate proportional to $m_b/t_{\mathrm{dyn}}$, where
$m_b$ is the mass of gas in the cells and $t_{\mathrm{dyn}}$ is the
local dynamical time.  These galaxy particles subsequently contribute
to metal production and the background ionizing UV radiation.  This
algorithm is essentially the same as that used by \cite{katz92a} and
Gnedin (1996a,b). 
The masses of these galaxy particles range from
about $10^6$ to $10^9$ $M_\odot$. Thus, many galaxy particles are
contained in what would correspond to a single luminous galaxy in the
real universe. However, rather than grouping the particles into
galaxies, in this paper we simply define a galaxy mass density field
from the distribution of galaxy particles themselves. More details
concerning the galaxy formation criteria and their consequences are
given in Appendix \ref{formation}.

We will examine the results of the simulations at four output times: $z=3$,
$z=1$, $z=0.5$ and $z=0$. At each output time, we will consider the properties
both of all the galaxy particles and of those galaxy particles formed within
the previous 0.5 Gyrs. We will take these ``recently'' formed particles to be
representative of the properties of galaxy formation at that output time, and
label their overdensity field $\delta_\ast\equiv\rho_\ast/\avg{\rho_\ast} - 1$,
the ``galaxy formation density field.''

%
%
\setcounter{section}{2}
\section{Time dependence of Galaxy Formation}
\label{timeformation}

In this section, we study the relationship between the mass density,
the gas temperature, and the galaxy formation density fields at each
epoch.  We will show the evolution of the the quantities $b_\ast$ and $r_\ast$ 
with redshift, and demonstrate that this evolution is largely due to
the changing relationship between mass density and temperature.

\begin{figure}
\figurenum{\fignum}
\epsscale{1.0}
\plotone{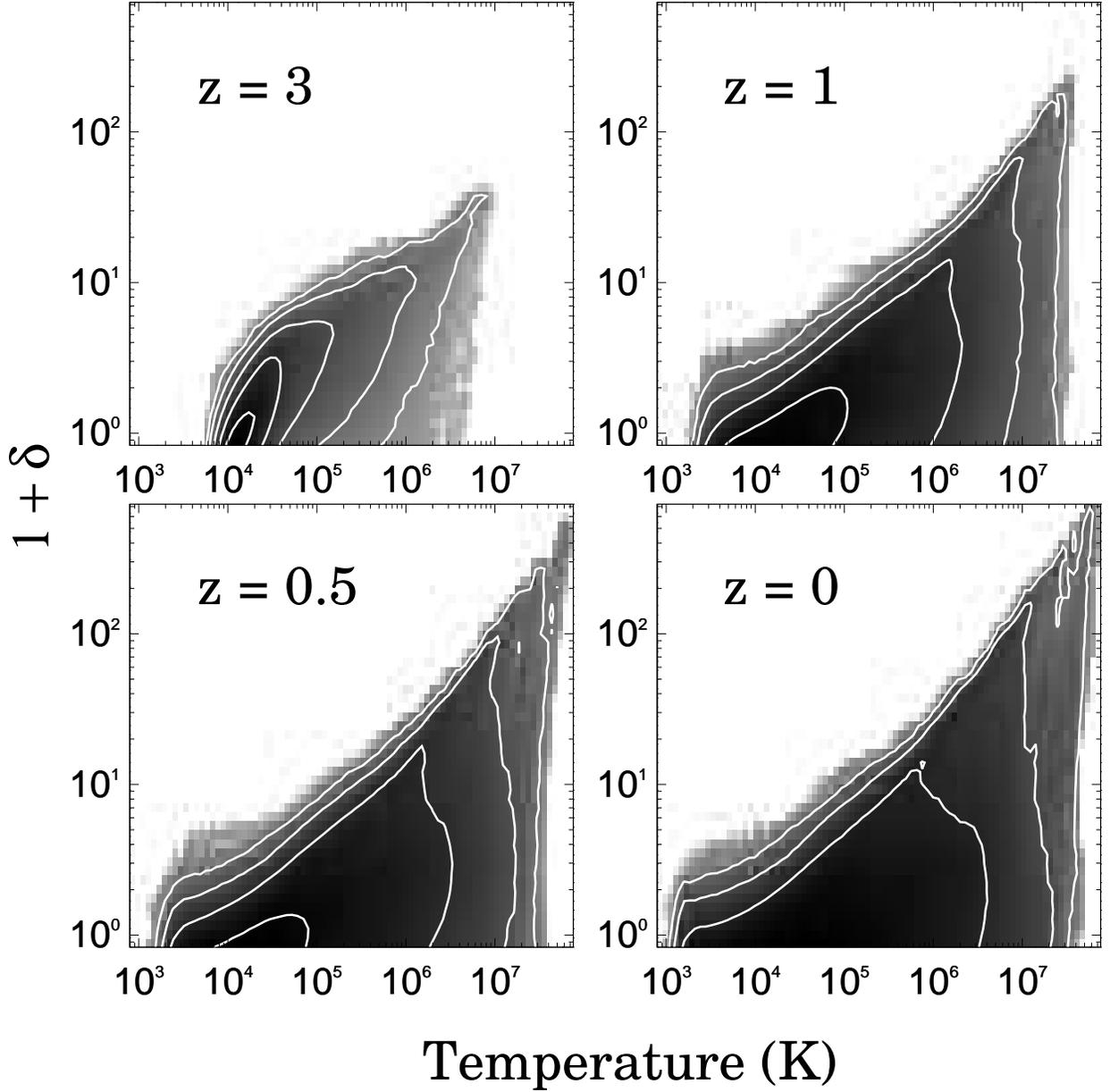}
\epsscale{1.0}
\caption{\label{dT} Volume-weighted joint probability distribution of
$1+\delta$, the mass density, and $T$, the gas temperature, for each output
redshift. Contours are spaced logarithmically. Each field has been smoothed
with a 1 $h^{-1}$ Mpc top hat. Note that the range of possible temperatures at
a fixed density increases with cosmic time. }
\end{figure}

Let us begin by considering the evolution of the temperature and mass density
fields as a function of redshift. At each of our four output times, we smooth
the fields with a top hat of 1 $h^{-1}$ Mpc comoving radius. In Figure \ref{dT}
we show the joint distribution of $1+\delta$, the normalized mass density, and
$T$, the gas temperature, as a function of redshift. We could instead have
considered the distribution of $1+\delta_b$, the baryonic density; 
the results would have been nearly identical, because the dark matter
and baryonic overdensities are closely related on these scales. Most
of the gas, that is, does not cool, fall to the central parts of
galaxies, and form stars, but instead follows the same distribution as
the dark matter (\cite{cen98bp}). As the universe evolves, the
variance in density increases. In addition, as gas falls into the
potential wells, it shocks and heats up to the virial temperature
necessary to support itself in the potential well, raising the average
temperature of gas in the universe. Thus, volume elements above the
mean density move up and to the right on this diagram. These results
are consistent with previous work with the same code (for instance,
\cite{cen93bp}) and with other codes (for a summary, see
\cite{kang94ap}). Note that on 1 $h^{-1}$ Mpc scales, the importance
of cooling physics to the gas temperature is not immediately apparent;
in Appendix A, we show this same figure at the scale of a single cell,
which shows quite clearly the effects of including the radiative
physics of the gas.

\begin{figure}
\figurenum{\fignum}
\epsscale{1.0}
\plotone{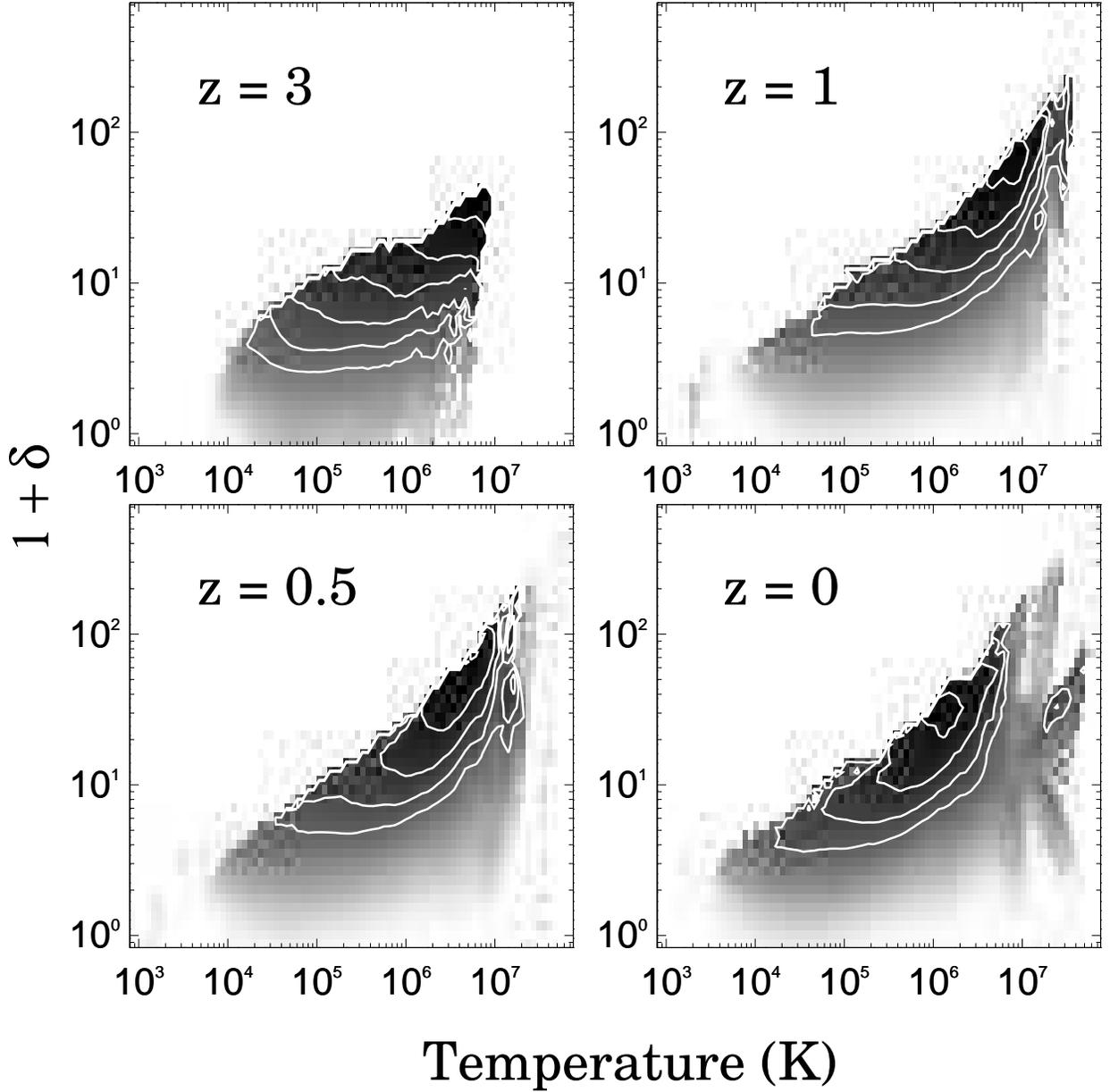}
\epsscale{1.0}
\caption{\label{dTg} The average rate of galaxy formation,
$1+\delta_\ast$, as a function of $1+\delta$ and $T$, for each output
redshift. Each field has been smoothed with a 1 $h^{-1}$ Mpc top
hat. We measure the local rate of galaxy formation by examining the
density distribution of ``recently-formed'' galaxy particles; that is,
those formed in the previous 0.5 Gyrs.  Contours are spaced
logarithmically. Note that at any given density, the star formation
declines with increasing temperature.  As the densest regions become
too hot to cool and collapse, they no longer become preferred sites
for galaxy formation, which move to the relatively less dense regions
of the universe. }
\end{figure}

\begin{figure}
\figurenum{\fignum}
\epsscale{1.0}
\plotone{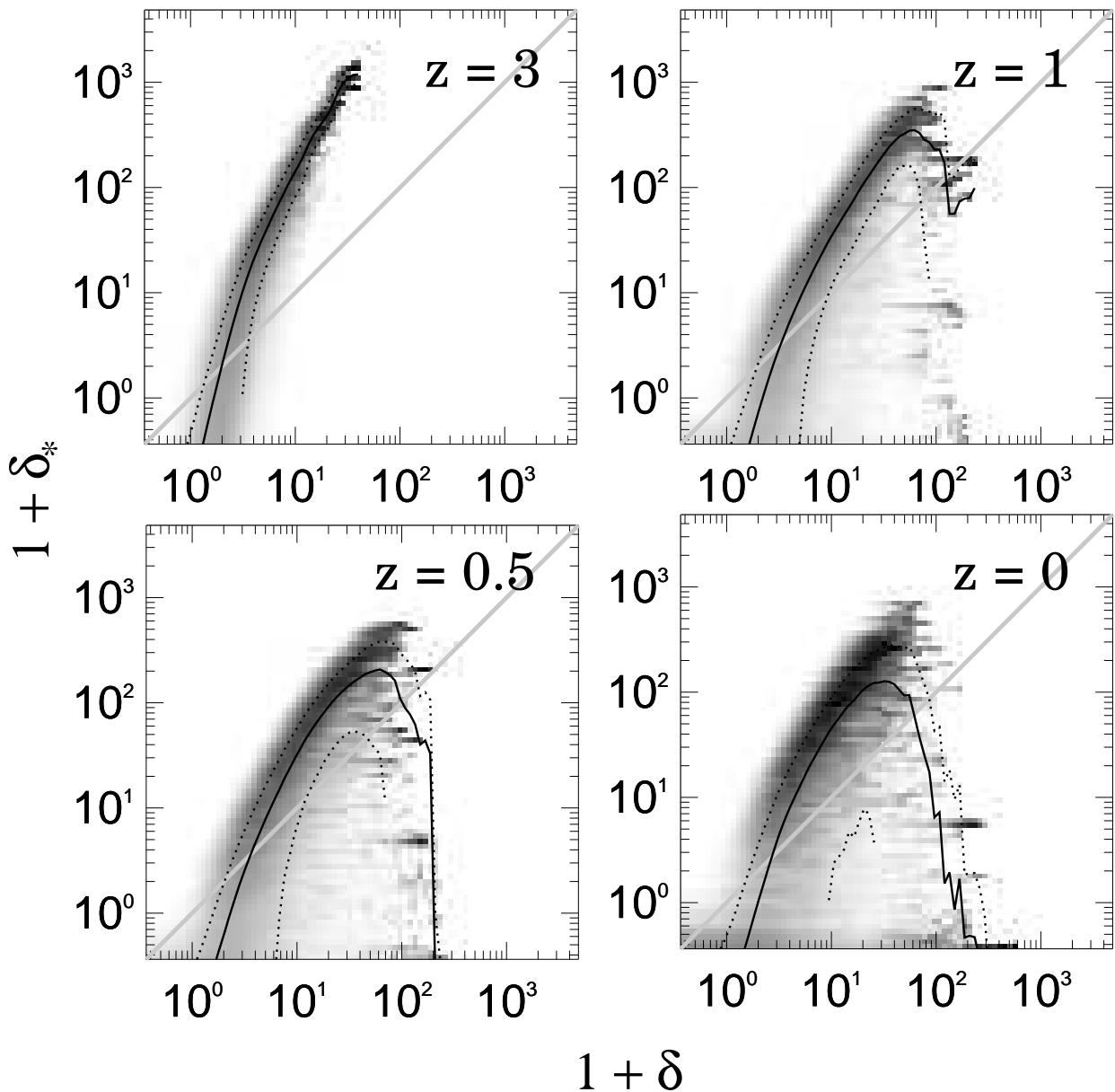}
\epsscale{1.0}
\caption{\label{dendenstar} Conditional probability $P(1+\delta_\ast|1+\delta)$
at each of four redshifts. Greyscale stretch is logarithmic. Each
field has been smoothed with a 1 $h^{-1}$ Mpc top hat. Solid line is
the conditional mean $\avg{1+\delta_\ast|1+\delta}$; dashed lines are
$1\sigma$ limits around that mean. Note that at later times there is
increasing stochasticity and a turnover in
$\avg{1+\delta_\ast|1+\delta}$ at high density. While in
\cite{blanton98a} we showed where galaxies of different ages ended up
at $z=0$, here we show instead where galaxies form at each redshift. }
\end{figure}

We now consider the equivalent plot for the galaxy formation density
field. Figure \ref{dTg} shows the conditional mean of the galaxy formation
density field $1+\delta_\ast$ as a function of $1+\delta$ and $T$:
$\avg{1+\delta_\ast|1+\delta, T}$.  
At any fixed density, the star formation rate declines as temperature
rises.
It is clear that the densest regions cannot cool and collapse at late
times because they are too hot.  To illustrate what Figure \ref{dTg}
implies about the relationship between galaxy formation and mass
density, Figure \ref{dendenstar} shows the conditional probability
$P(1+\delta_\ast|1+\delta)$ as the logarithmic grey scale and the
conditional mean $\avg{1+\delta_\ast|1+\delta}$ as the solid line. The
dashed lines are 1$\sigma$ limits around the mean. At high redshift,
the high density regions are still cool enough to have clumps of gas
in them cooling quickly; in this sense, the temperature on these
scales is not particularly important, making the relationship between
galaxy formation and mass density close to deterministic. At low
redshift, many regions become too hot to form new galaxies; in
particular, galaxy formation is completely quenched in the densest
regions, causing a turnover in $\avg{1+\delta_\ast|1+\delta}$. In
addition, at low redshifts, the galaxy formation rate depends strongly
on temperature, adding scatter to the relationship between galaxy
formation and mass (\cite{blanton98ap}).

\begin{figure}
\figurenum{\fignum}
\epsscale{1.0}
\plotone{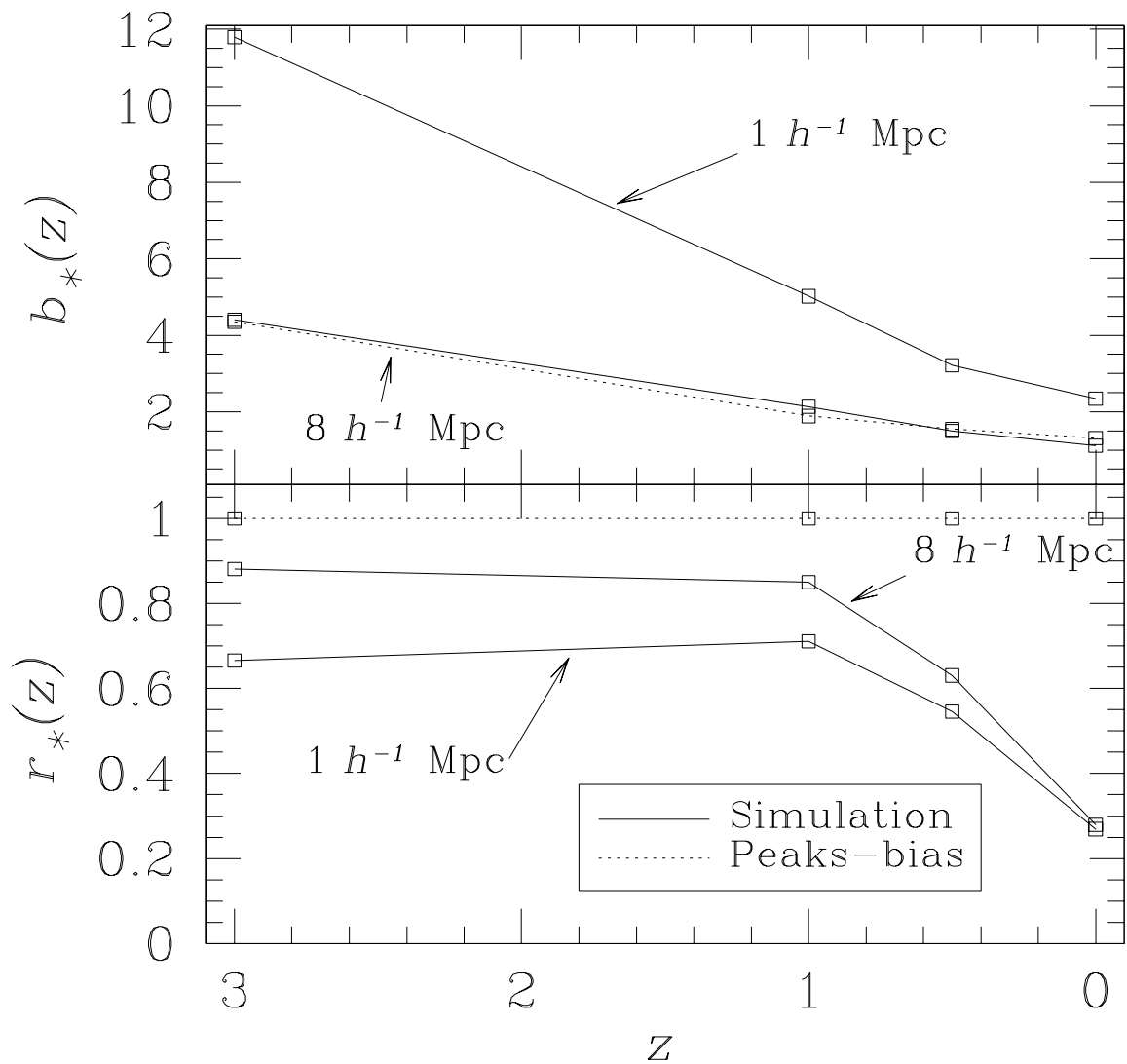}
\epsscale{1.0}
\caption{\label{brstartime} The redshift evolution of the bias
$b_\ast(z)\equiv\sigma_\ast/\sigma$ (top) and the correlation
coefficient $r_\ast(z)\equiv\avg{\delta\delta_\ast}/\sigma\sigma_\ast$
(bottom) ``at birth,'' for top hat smoothing scales of 1 $h^{-1}$ Mpc
and 8 $h^{-1}$ Mpc. There is a strong trend with redshift. We also
show, as the dotted lines, the peaks-bias prediction for $b_\ast(z)$
and $r_\ast(z)$ on large scales, for $M>10^{12} M_\odot$. Note that
$r_\ast(z)$ does not evolve at all in the peaks-biasing model.}
\end{figure}

The most obvious consequence of this evolution is that the bias of
galaxy formation is a strong function of time.  Figure
\ref{brstartime} shows $b_\ast$ and $r_\ast$ (defined in Section
\ref{motiv}) as a function of redshift for top hat smoothing radii of
$R=1$ and $R=8$ $h^{-1}$ Mpc. The bias of galaxy formation $b_\ast$ is
clearly a strong function of redshift, especially at small scales. The
decrease in the bias is due to the fact that the galaxy formation has
moved to lower $\sigma$ peaks and out of the hottest (and densest)
regions of the universe, as described in Section \ref{motiv}.

We can compare these results to those of the peaks-bias formalism for
the bias of dark matter haloes.  If we assume that the distribution of
galaxy formation is traced by $M > 10^{12}$ $M_\odot$ halos (that is,
halos with typical bright galaxy masses) which have just collapsed,
then according to the peaks-bias formalism (\cite{doroshkevich70ap};
\cite{kaiser84ap}; \cite{bardeen86ap}; \cite{bond91ap};
\cite{mo96ap}), the bias of mass in such halos with respect to the
general mass density field is:
\begin{equation}
b_\ast(z) = 1 + \frac{\delta_c}{\sigma^2(M,z)} \mathrm{,}
\end{equation}
where $\delta_c\sim 1.7$ is the linear overdensity of a sphere which
collapses exactly at the redshift of observation, and $\sigma(M,z)$ is
the rms mass fluctuation on scales corresponding to mass $M$, again at
the redshift of the observations.  Note that this formula differs from
the standard one for the bias of all halos which have collapsed before
redshift $z$. The dotted line in the upper panel of Figure
\ref{brstartime} shows the resulting bias. Thus, the prediction for
$b_\ast$ of the pure halo bias model is not much different from that
of our hydrodynamic model on large scales, even though the former does
not include all the physical effects of the latter.
That is, measuring
$b_\ast(z)$ alone cannot easily distinguish the two models.

More interesting is the evolution of the correlation coefficient
$r_\ast$ as a function of redshift, shown in the lower panel of Figure
\ref{brstartime}. At all scales, the correlation coefficient falls
precipitously for $z<1$, when the densest regions become too hot to
form new galaxies.  On the other hand, in the case of peaks-biasing,
\cite{mo96a} find a scaling between the halo and mass cross- and
autocorrelations that implies $r_\ast\sim 1$ on large scales at all
redshifts\footnote{On scales of 1 $h^{-1}$ Mpc, things are more
complicated, because $10^{12}$ $M_\odot$ halos so close together
actually overlap in Lagrangian space --- see \cite{sheth98ap}.}. This
is an important difference between the predictions of the models of
\cite{mo96a} and ours, which is presumably due to the more realistic
physics used in our model. Semi-analytic models can predict the
evolution of $r_\ast$; it will be interesting to see how dependent
this statistic is to the details of the galaxy formation model.  We
discuss a way to get a handle on $r_\ast$ observationally in Section
\ref{observables}.



%
%
\setcounter{section}{3}
\section{Gravitational Evolution of Bias}
\label{burst}

After galaxies form, they fall into potential wells under the influence of
gravity. Because the acceleration on galaxies is the same as that on the dark
matter, this gravitational evolution after formation will tend to bring both
the bias $b_g$ and the correlation coefficient $r_g$ closer to unity, as
described by \cite{fry96a} and \cite{tegmark98a}.  The evolution of the bias,
then, is determined by the properties of forming galaxies (outlined in the
previous section) and how those properties evolve after formation. Here we
investigate this process and show that the linear approximations of
\cite{fry96a} and \cite{tegmark98a} describe this evolution well, even in the
nonlinear regime.

\subsection{Evolution of the Clustering of Coeval Galaxies}

\cite{fry96a} and \cite{tegmark98a} rely on the continuity equation to
study the evolution of the density field of galaxies formed at a given
epoch.  In this subsection, subscript $c$ will refer to this coeval
set of galaxies. Both galaxies and mass satisfy the same equation,
which is to first order 
\begin{equation}
\label{continuity}
{\dot \delta} + \nabla\cdot\vv{v} = 0\mathrm{,}
\end{equation}
assuming that galaxies are neither created nor destroyed. Since we are
dealing with a coeval set of galaxies, they cannot be created by
definition. Although it is possible to disrupt or merge galaxies in
dense regions (\cite{kravtsov98ap}), we are following the stellar mass
density in this paper, not the galaxy number density, and thus do not
consider these effects.

\begin{figure}
\figurenum{\fignum}
\epsscale{1.0}
\plotone{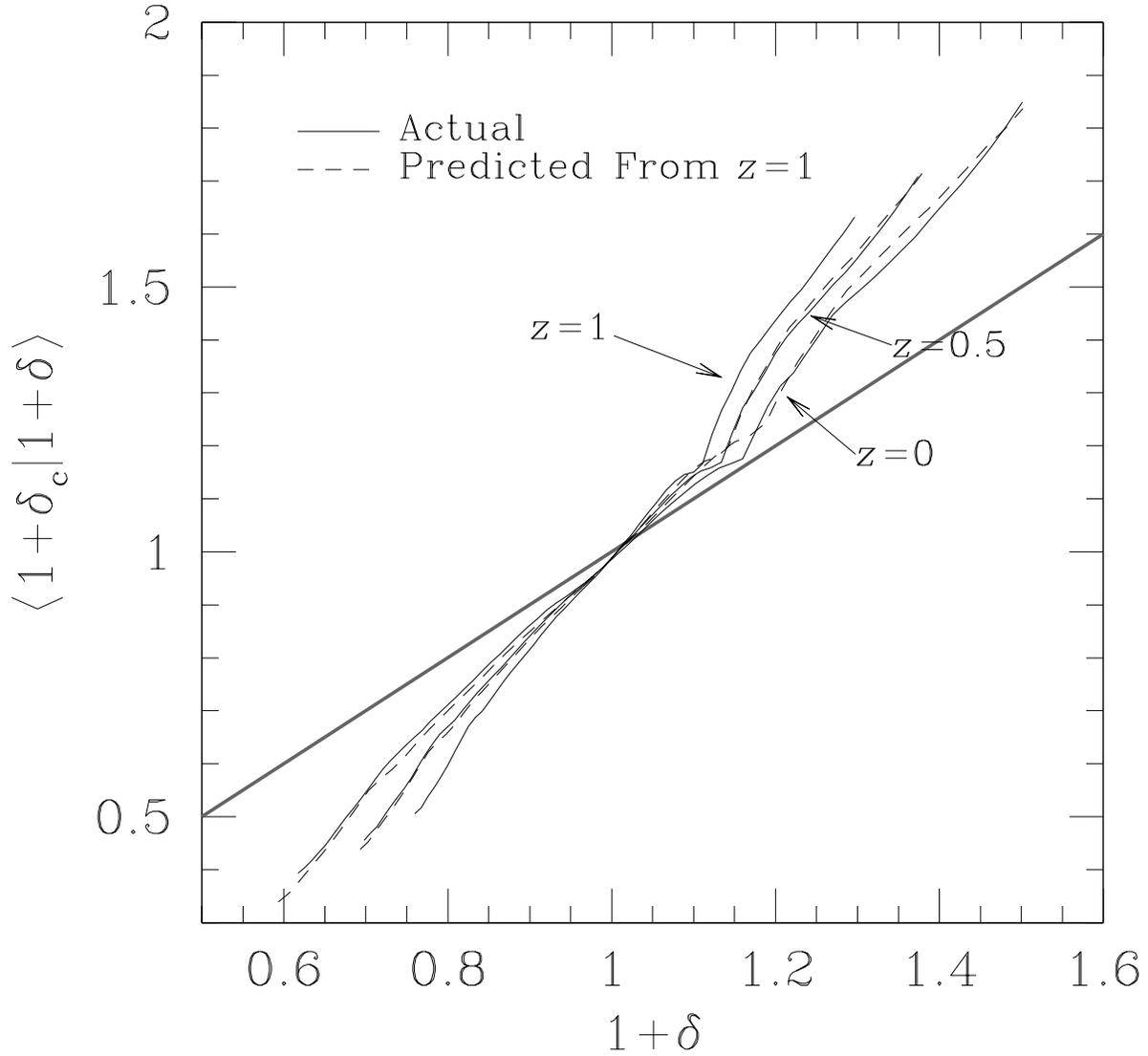}
\epsscale{1.0}
\caption{\label{evolve30} Evolution of a burst of galaxies at
$z=1$, smoothed on 30 $h^{-1}$ Mpc scales. 
The solid lines show the conditional mean galaxy density given
the mass density, at $z=1$, $z=0.5$ and $z=0$. The dashed lines show
the predictions at $z=0.5$ and $z=0$ given the results at $z=1$,
according to the rule that $\Delta\delta_c = \Delta\delta$. }
\end{figure}

\begin{figure}
\figurenum{\fignum}
\epsscale{1.0}
\plotone{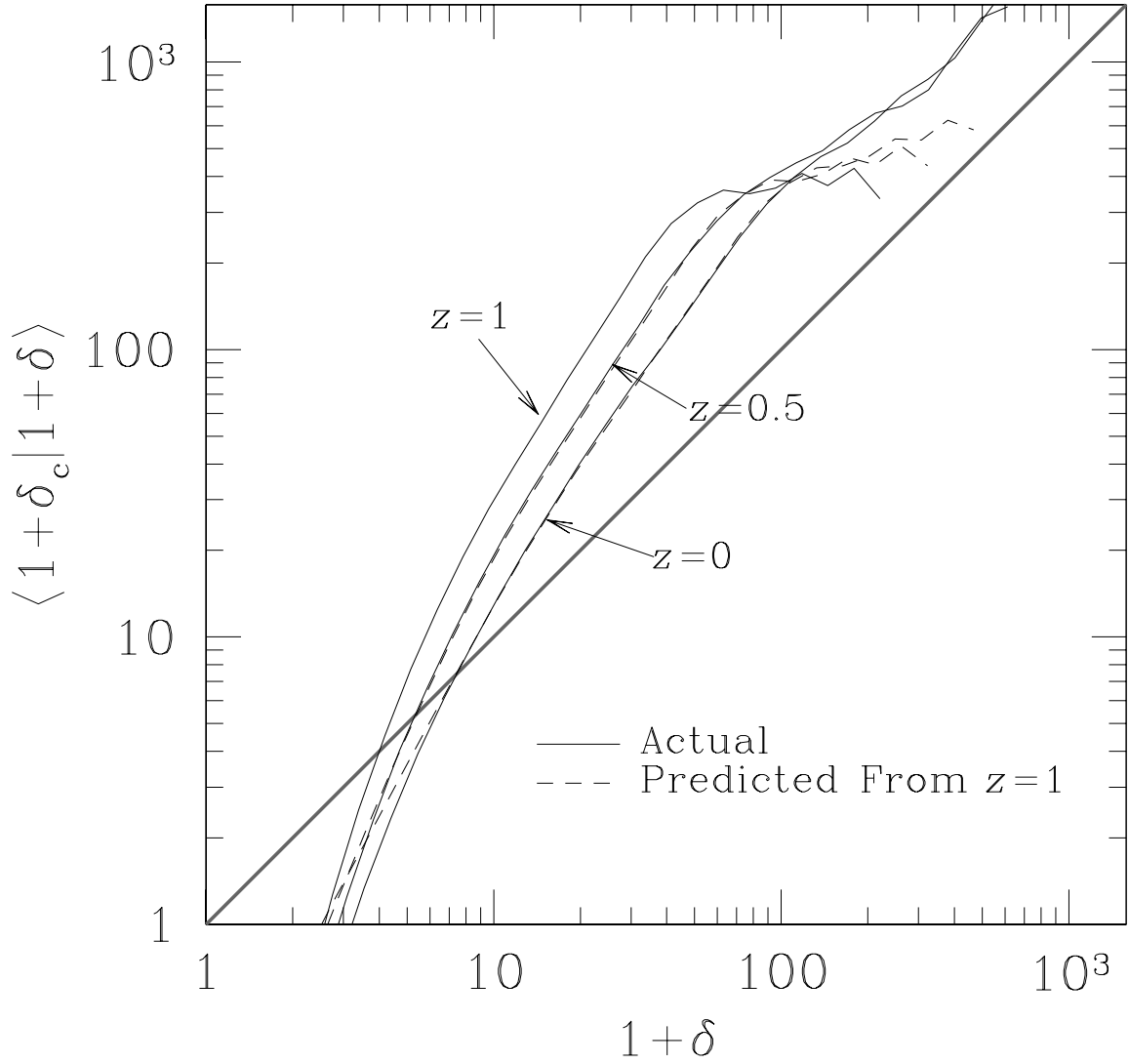}
\epsscale{1.0}
\caption{\label{evolve1} Same as Figure \ref{evolve30}, now smoothed
on 1 $h^{-1}$ Mpc scales and plotted on a logarithmic
scale. Predictions are good in the regime: $3 < \delta < 100$. }
\end{figure}

Assuming that there is no velocity bias between the volume-weighted velocity
fields of mass and of galaxies (which is borne out by the simulations), it
follows that ${\dot \delta} = {\dot \delta_c}$. This statement means that on
linearly scaled plots of $\delta_c$ against $\delta$, volume elements are
constrained to evolve along $45^\circ$ lines. We test this prediction on 30
$h^{-1}$ Mpc top hat scales, which should be close to linear, in Figure
\ref{evolve30}. Here we consider a burst of galaxies formed at $z=1$. The solid
lines are $\avg{\delta_c|\delta}$ at $z=1$, $z=0.5$ and $z=0$. The dashed lines
are the predictions extrapolated from $z=1$ as above, to
$z=0.5$ and $z=0$, respectively. Notice
use the growth factor of mass, $\sigma(z=0)/\sigma(z=1)$, which on these scales
is close to its linear theory value, to determine $\Delta\delta$, and assume
that $\Delta\delta_c$ is the same. Evidently the linear continuity equation
works on these scales. We also test this prediction on the nonlinear scale of 1
$h^{-1}$ Mpc in Figure \ref{evolve1}. We find that it works extremely well in
the range $3 < \delta < 100$; in higher and lower density regions, nonlinear
corrections to Equation (\ref{continuity}) are clearly important.

\begin{figure}
\figurenum{\fignum}
\epsscale{1.0}
\plotone{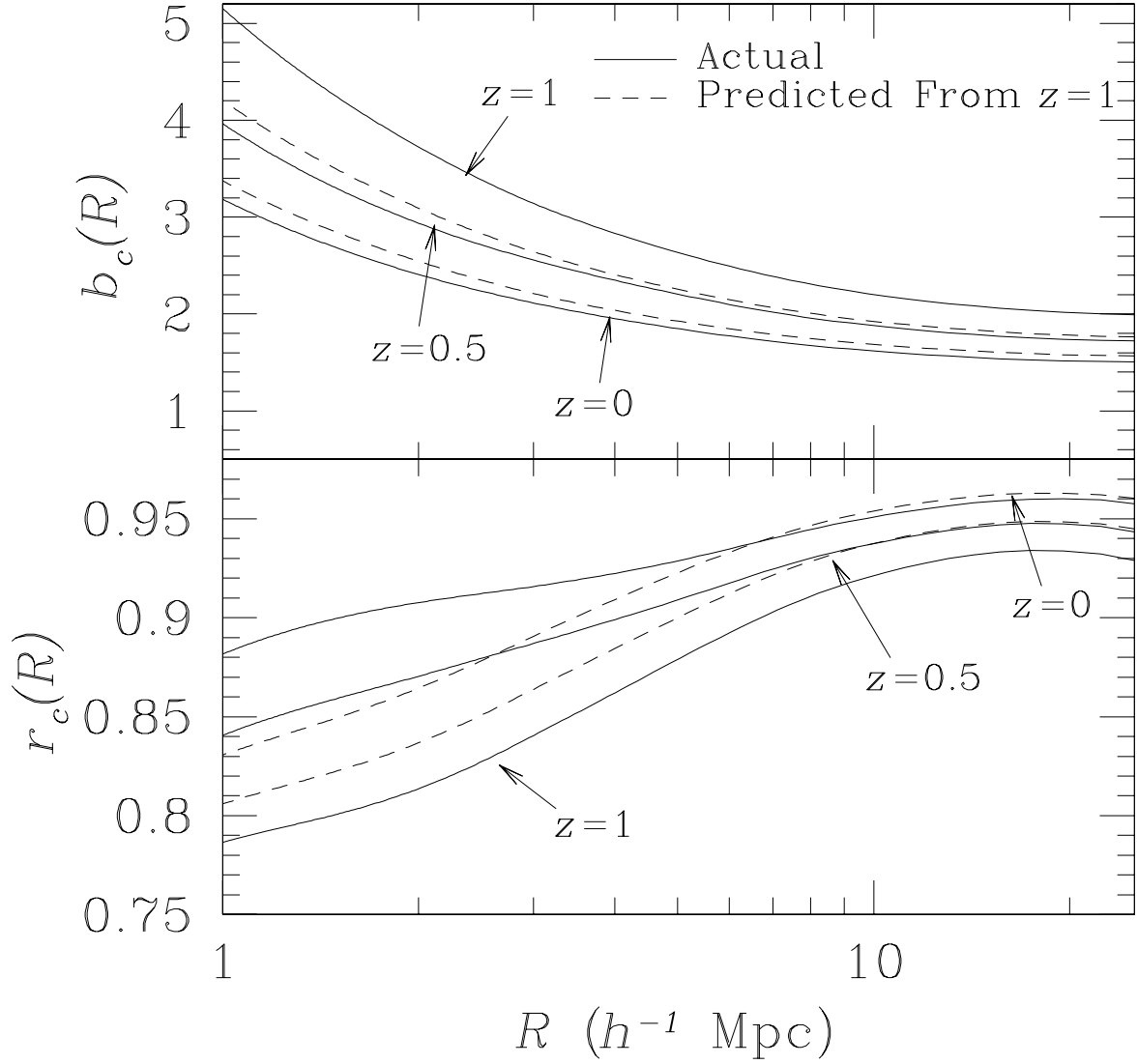}
\epsscale{1.0}
\caption{\label{brevolve} Evolution of the bias $b_c(R)$ and correlation
coefficient $r_c(R)$ of a burst of galaxies at $z=1$. Solid and dashed
lines have the same significance as in Figure \ref{evolve30}. In this
case, predictions use Equation (\ref{bandr}). }
\end{figure}

\cite{tegmark98a} use the continuity equation to make further
predictions about the bias properties of coeval galaxies. In
particular, given a ``bias at birth'' of $b_c(z_0)$ and a
``correlation coefficient at birth'' of $r_c(z_0)$, one can express
$b_c(z)$ and $r_c(z)$ in terms of the linear growth factor relative to
the epoch of birth $D(z)/D_0$:
\begin{eqnarray}
\label{bandr}
b_c(z)r_c(z) &=& 1 + \frac{b_c(z_0) r_c(z_0) - 1}{D(z)/D_0}\cr
b_c^2(z) &=& \frac{\left[(1-D(z)/D_0)^2 - 2(1-D(z)/D_0)b_c(z_0) r_c(z_0) +
b_c^2(z_0)\right]}{(D(z)/D_0)^2}
\end{eqnarray}
In Figure \ref{brevolve} we show as the solid lines the values of $b_c$ and
$r_c$ for the burst of galaxies at $z=1$, at each output time. The dashed lines
are the predictions based on $z=1$ for the results at $z=0.5$ and $z=0$. Notice
that on small scales, the prediction for $b_c$ remains extremely good. For
$r_c$, the linear prediction underestimates the degree to which galaxies and
mass become correlated on small scales; essentially, nonlinearities enhance the
rate at which initial conditions are forgotten.

\begin{figure}
\figurenum{\fignum}
\epsscale{1.0}
\plotone{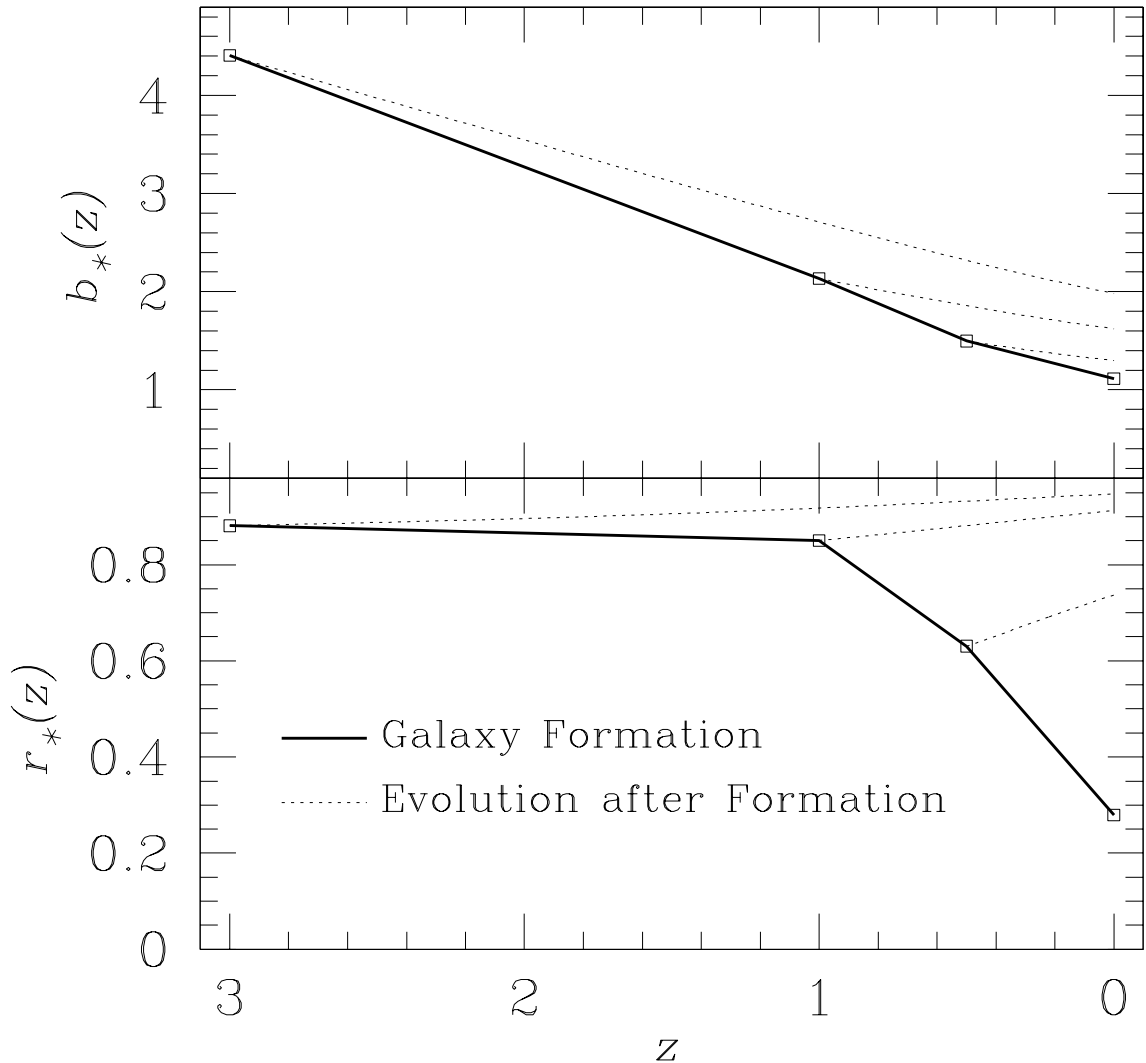}
\epsscale{1.0}
\caption{\label{debias} A comparison of the debiasing due to gravitational
evolution and due to the shift in the location of galaxy formation. The solid
line shows the evolution of galaxy formation, smoothed with an 8 $h^{-1}$ Mpc
radius top hat, from Figure \ref{brstartime}. 
The dotted lines emanating from each output time indicate how
the bias properties of galaxies formed at that time evolve according to the
continuity equation. The top panel shows
that gravitational debiasing acts more slowly than the shift of galaxy
formation out of the high density regions. This is why old galaxies at redshift
zero are still more highly biased than young galaxies, despite having had more
time to debias. The bottom panel shows how the decline in $r_\ast$ tends to be
washed out by the debiasing. }
\end{figure}

We can compare the effects of debiasing to the decline in the
correlation between galaxy formation and mass with redshift. In Figure
\ref{debias} we show $b_\ast(z)$ and $r_\ast(z)$ at 8 $h^{-1}$ Mpc
scales, from Figure \ref{brstartime}, as the solid lines. The dotted
lines originating at each output redshift show how the galaxies formed
at that redshift would evolve according the continuity
equation. Naturally, $b_c(z)$ declines while $r_c(z)$ rises. Note that
$b_\ast(z)$ falls more quickly than debiasing can occur, which is why
old galaxies at redshift zero are still more highly biased than young
galaxies (\cite{blanton98ap}), despite having had more time to debias.

\subsection{Evolution of the Clustering of the Full Galaxy Population}

\begin{figure}
\figurenum{\fignum}
\epsscale{1.0}
\plotone{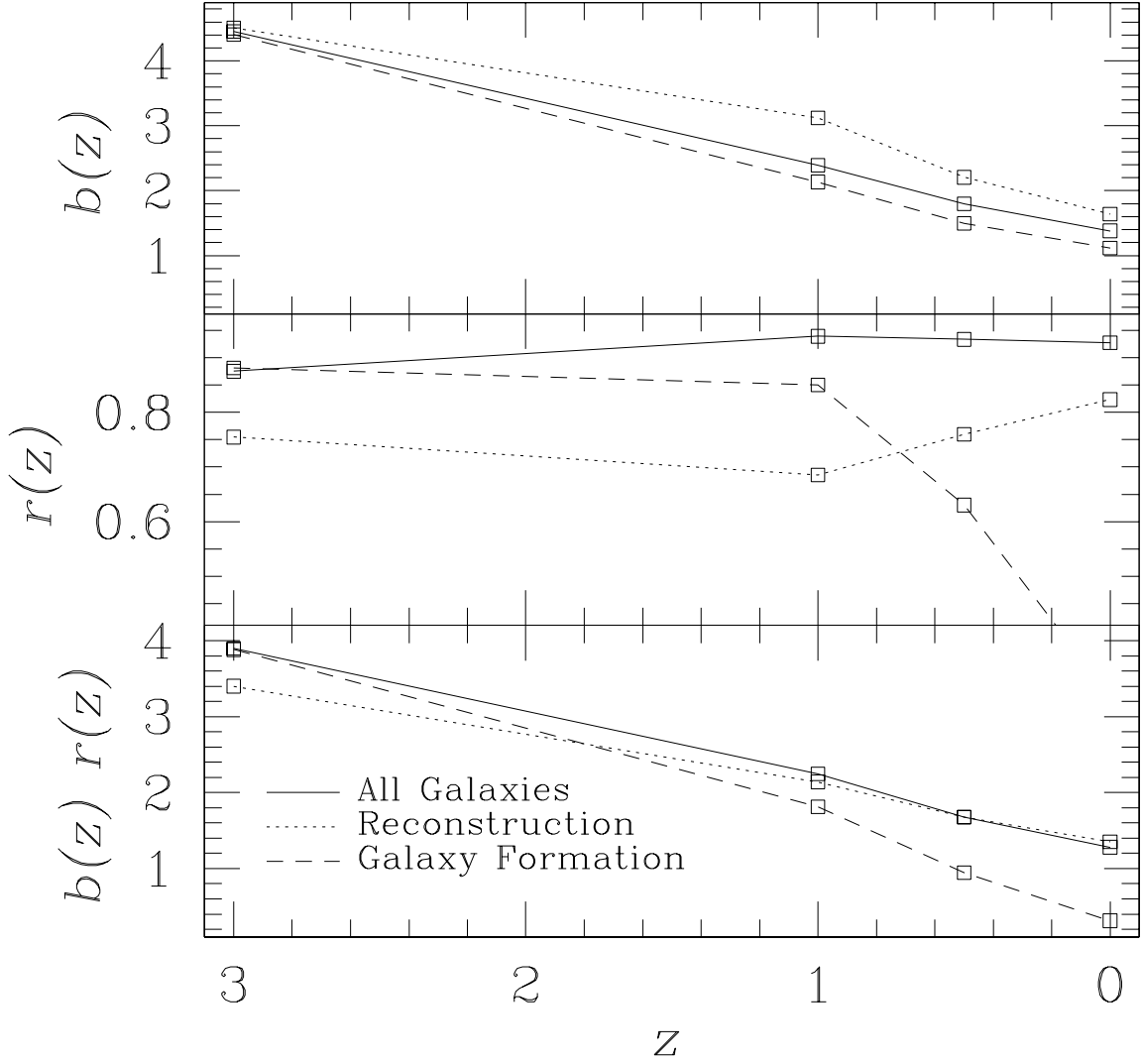}
\epsscale{1.0}
\caption{\label{csstartime} Evolution of the bias $b_g(z)$, correlation
coefficient $r_g(z)$, and the combination $b_g(z)r_g(z)$ of all galaxies on
8 $h^{-1}$ Mpc tophat scales. Solid line is measured from the
simulations. Dotted line is the reconstruction according to 
Equation (\ref{fullev}). The reconstruction suffers from the poor time
sampling of our output. The dashed lines are the corresponding
quantities for the recently-formed galaxies at each redshift, as in Figure
\ref{brstartime}. }
\end{figure}

These results show that the decrease in the bias of the galaxy density field
with time found by \cite{cen98a} must be partly due to gravitational debiasing
(described in this section) as well as the decrease in $b_\ast$ (described in
Section \ref{timeformation}). We investigate this process here by showing the
evolution of the bias of all the galaxies in Figure \ref{csstartime} as the
solid lines, for a tophat smoothing of 8 $h^{-1}$ Mpc. The evolution of the
bias of galaxy formation, from Figure \ref{brstartime}, is shown as the dashed
lines. Notice that $b_g(z)$ and $b_\ast(z)$ are nearly the same, even though
$r_g(z)$ and $r_\ast(z)$ differ considerably; $r_\ast(z)$ plummets after $z=1$,
while $r_g(z)$ remains close to unity. About half the star formation in the
simulation occurs at $z<1$.  Although this recent star formation has a
low $r_\ast$, and is thus poorly
correlated with the mass distribution, this has little effect on
$r_g$, partly due to the gravitational debiasing.

Using the approximations of \cite{tegmark98a}, we can reconstruct
$b_g(z)$ and $r_g(z)$ from the properties of galaxy formation as a
function of redshift. We get equations similar to those of
equation~(\ref{bandr}), now integrated over the star formation
history: 
\begin{eqnarray}
\label{fullev}
b_g(z) r_g(z) &=& 1 + \frac{1}{D(z) G(z)} \int dz' \left[b_\ast(z') r_\ast(z')
- 1\right] D(z') \frac{dG}{dz'}\mathrm{,}\cr b_g^2(z) &=& b_g^2(z) r_g^2(z) +
\frac{s^2}{\sigma^2}\mathrm{,}
\end{eqnarray}
where $D(z)$ is the linear growth factor, $G(z)$ is the cumulative
mass of stars formed by redshift $z$, and $\sigma$ is the rms mass
fluctuation. Calculating $b_g(z)$ requires first evaluating
$b_g(z)r_g(z)$ substituting it into the expression given.  In the
second term, the quantity $s^2$ is defined as:
\begin{equation}
s^2 \equiv \frac{1}{G^2(z)} \int dz' \frac{dG'}{dz'} \int dG''
\frac{dG''}{dz''} \avg{\delta_\perp(z') \delta_\perp(z'')} \mathrm{,}
\end{equation}
where $\delta_\perp \equiv \delta_\ast - b_\ast r_\ast \delta$ is the component
of the galaxy formation field which is uncorrelated with the local
density.\footnote{Note that the definition of $b_\ast$ in \cite{tegmark98a} is
equivalent to our quantity $b_\ast r_\ast$.}  These equations are simply a
result of applying the continuity equation (Equation \ref{continuity}) to the
case of continuous star formation, rather than an instantaneous burst, as in
Equation (\ref{bandr}).  Figure \ref{csstartime} shows $b_g(z)$, $r_g(z)$, and
$b_g(z) r_g(z)$ for all galaxies reconstructed from the properties of our
galaxy formation fields. $b_g(z)r_g(z)$ represents the linear regression of the
galaxy density on the mass density. Note that the reconstruction of this
quantity is quite accurate; on the other hand, the reconstructions of $b_g(z)$
and $r_g(z)$ separately, which require knowledge of $\avg{\delta_\perp(z')
\delta_\perp(z'')}$, have significant errors. 
These errors are
mostly due to the poor time resolution of our output.\footnote{The results of
these simulations take large amounts of disk space, and thus when they were run
we did not attempt to store more than a handful of output times. It would be
impractical to rerun the simulations again for the sole purpose of obtaining
more output redshifts.} The continuity equation will prove
useful to future theoretical and observational work, as we discuss in Section
\ref{conclusions}.

%
%
\setcounter{section}{4}
\section{Observing the Evolution of Galaxy Formation}
\label{observables}

Here we discuss the observational consequences of the evolution of the spatial
distribution of galaxy formation.  First, we consider
observable properties of the star formation density field.  Second, we consider
the resulting properties of galaxy clusters.

\subsection{Star Formation Density Field}

\begin{figure}
\figurenum{\fignum}
\epsscale{1.0}
\plotone{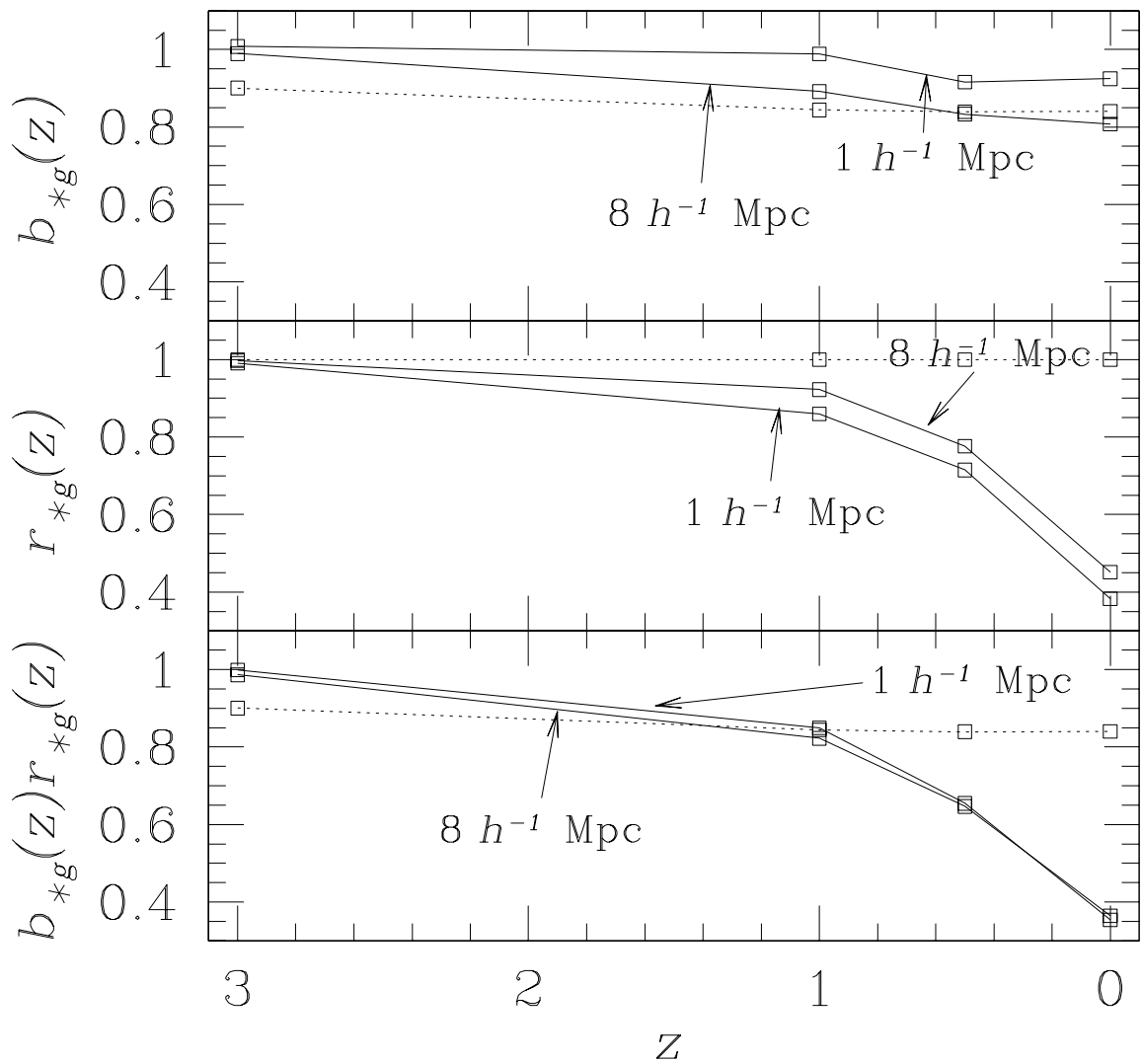}
\epsscale{1.0}
\caption{\label{grecent} Bias $b_{\ast g}(z)\equiv\sigma_\ast/\sigma_g$,
correlation coefficient $r_{\ast
g}(z)\equiv\avg{\delta_\ast\delta_g}/\sigma_\ast\sigma_g$, and the linear
regression $b_{\ast g} r_{\ast g}$ of the galaxy formation density field on
the density field of all galaxies as a function of redshift. We show
results for 1 $h^{-1}$ Mpc and 8 $h^{-1}$ Mpc radius tophat
smoothing. The dotted lines are the peaks-bias predictions 
for the same quantities on large scales, comparing the bias of the
mass in recently formed halos $> 10^{12} M_\odot$ to that of all halos
$> 10^{12} M_\odot$.}
\end{figure}

We noticed in the last section that the correlation coefficient
between star forming galaxies and mass should decrease considerably
with time. We cannot currently observe this decrease directly in the
real universe.\footnote{Weak lensing studies (\cite{mellier99ap}) and
peculiar velocity surveys (\cite{strauss95ap}) do probe the mass
distribution and may prove useful for studying the evolution of
$b_g(z)$ and $r_g(z)$ directly.} However, from Figure \ref{csstartime} we note
that the galaxy distribution as a whole {\it does} correlate well with
the mass, in the sense that $r_g(z)\sim 1$ at all
redshifts. Therefore, we should be able to detect the evolution of
$r_\ast(z)$ by cross-correlating galaxy formation with the
distribution of all galaxies. In Figure \ref{grecent} we plot $b_{\ast
g}\equiv\sigma_\ast/\sigma_g$ and $r_{\ast
g}\equiv\avg{\delta_\ast\delta_g}/\sigma_\ast\sigma_g$ between the
galaxy formation density field and all galaxies as a function of
redshift, at 8 $h^{-1}$ Mpc scales. While the evolution in $b_{\ast
g}$ is rather weak, the evolution in $r_{\ast g}$ is striking, and
should be observable.

Observationally, measuring this correlation will require mapping the
density field of {\it star formation} in the universe, which has not
yet been done. Given that the fundamental prediction of galaxy
formation models is the location of star formation as a function of
time, such a map would be extremely useful. The spectral coverage of
the Las Campanas Redshift Survey (LCRS) includes a star formation
indicator ([OII] equivalent widths; \cite{schechtman96ap};
\cite{hashimoto98ap}); thus, one can measure this correlation at low
redshift using this data.  Using the LCRS, \cite{tegmark98b} have
addressed a related question by measuring the correlation between the
distributions of early and late spectral types (as classified by
\cite{bromley98ap}); they found $r\sim 0.4$--$0.7$, in qualitative
agreement with our findings here, if one makes a correspondence
between spectral type and star formation rate.  They can calculate $r$
because they understand the properties of their errors, which are due
to Poisson statistics; however, in the case of measuring spectral
lines to determine the star formation rate in a galaxy, the errors are
much greater and more poorly understood. Therefore, since $\delta_g$
is likely to be much better determined than $\delta_\ast$, the
quantity of interest will not be $r_{\ast g}$, but the combination
$b_{\ast g} r_{\ast g}=\avg{\delta_\ast\delta_g}/\sigma_g^2$, which is
the linear regression of star formation density on galaxy density and
is independent of $\sigma_\ast$.  Future redshift surveys such as the
SDSS will probe redshifts as high as $z\sim 0.2$---$0.3$ and will have
spectral coverage which will include measures of star formation such
as [OII] and H$\alpha$. Figure \ref{csstartime} indicates that $b_{\ast
g} r_{\ast g} \sim 0.35$ at $z=0$ and $b_{\ast g} r_{\ast g} \sim 0.5$
at $z=0.25$ (linearly interpolating between $z=0$ and $z=0.5$), a
difference that should be measureable.  A cautionary note is that the
linear interpolation in this range 
may not be accurate, and that simulations with finer time
resolution are necessary to make more solid predictions.

In addition, the SDSS photometric survey will probe to even higher
redshift. Using the multiband photometry, one will be able to produce
estimates both of redshift and of spectral type. These measurements
will allow a calculation of the angular cross-correlation of galaxies
of different types or colors as a function of redshift out to
$z\sim 0.5$---$1$; Figure \ref{grecent} shows that this
evolution should be strong. The drawback to this approach is that one
cannot make a quantitative comparison between our models and the
cross-correlations of galaxies of different colors, because we can not
yet reliably identify individual galaxies in the simulations and
because there is not a one-to-one correspondence between galaxy color
and ages.  Needless to say, the observational question is still
worth asking, because quantitative comparisons will be possible in the
future using hydrodynamical simulations at higher resolution, and are
even possible today using semi-analytic models (\cite{kauffmann98ap};
\cite{somerville98ap}; \cite{baugh98ap}).


\subsection{Clusters of Galaxies}

The measurement of $b_{\ast g}$ and $r_{\ast g}$ directly in the field
is not possible given current data. However, the observed evolution of galaxy
properties in clusters
does give us a handle on the evolution
of the distribution of galaxy formation. Here, we investigate cluster
properties in the simulations. First, we describe the effect of the
clustering of galaxy formation on cluster mass-to-light ratios, and
how that subsequently affects estimates of $\Omega$ based on
mass-to-light ratios. Second, we discuss the Butcher-Oemler (1978,
1984) effect.
We select the three objects in the simulation with the largest
velocity dispersions, whose properties are listed in Tables
\ref{clustersa} and \ref{clustersb} at redshifts $z=0.5$ and
$z=0$. This number of clusters of such masses in the volume of the
simulations is roughly consistent with observations
(\cite{bahcall98ap}). Clusters \#2 and \#3 are about 5 $h^{-1}$ Mpc
away from each other at $z=0$, and have a relative velocity of 1000 km
s$^{-1}$ almost directly towards each other; thus, they will merge in
the next 5 Gyrs.

To calculate mass-to-light ratios, we must calculate the luminosity associated
with each galaxy particle. There are two steps: first, we assign a model for
the star formation history to each particle; then, given that star formation
history, we use spectral synthesis results provided by Bruzual \& Charlot
(1993, 1998) to follow the evolution of their
rest-frame luminosities in $B$, $V$, and $K$. The simplest model for the
star formation history of a single particle is that its entire mass forms stars
immediately, the instantaneous burst (IB) approximation. 
However, it is unrealistic to expect that all of the star formation
will occur immediately; at minimum, it will at least a dynamical time
for gas to collapse from the size of a resolution element to the size
of a galaxy. Thus, we also consider an alternative model based on the
work of Eggen, Lynden-Bell, \& Sandage (1962; hereafter ELS), which
spreads the star formation out over the local dynamical time.  We
assign a star formation rate of
\begin{equation}
\frac{dm_\ast}{dt} = \frac{m_p}{t_{\mathrm{dyn}}} \frac{t}{t_{\mathrm{dyn}}}
e^{-t/t_{\mathrm{dyn}}},
\end{equation}
to each particle, where $m_p$ is the mass of the galaxy particle and
$t_{\mathrm{dyn}}$ is the local dynamical time at the location of the
particle at the time it was formed. 
The evolution of the mass-to-light ratios of clusters
are qualitatively the same for both the IB and ELS models; the
evolution of the $B-V$ colors of cluster galaxies does differ
somewhat between the models, as described below.

We can calculate the mass-to-light ratios of these clusters in the
$B$-band, $\Upsilon_B \equiv (M/L_B)/(M_\odot/L_{\odot, B})$, and
compare them to the critical mass-to-light ratio $\Upsilon_{B,c}$
necessary to close the universe. We do the same in the $V$ and
$K$-bands, and show the results in Tables \ref{clustersa} and
\ref{clustersb} at redshifts $z=0$ and $z=0.5$, respectively.  An
observer in the simulated universe would thus estimate
$\Omega_{\mathrm{est}}(z=0) \equiv \Upsilon_{B}/\Upsilon_{B,c}\approx
0.4$--$0.5$ based on cluster mass-to-light ratios, not far from the
correct value for the simulation ($\Omega_0=0.37$).  The galaxy bias,
which causes $\Omega_{0,\mathrm{est}}$ to underestimate $\Omega_0$ by
a factor of about 1.6, is partially cancelled by the relative fading
of the light of the older cluster galaxies with respect to the younger
field galaxies.
The bias effect (as we have shown above) increases in
importance with redshift, while the differential fading of cluster
galaxies decreases in
importance. Thus, in the simulations, the cluster mass-to-light ratio
estimates of $\Omega_{\mathrm{est}}(z)$ decrease with redshift, even
though $\Omega(z)$ itself increases. We find similar results for 
analyses performed in $V$ and $K$, which we also list in Tables
\ref{clustersa} and \ref{clustersb}. For the ELS star formation model,
the mass-to-light ratios are all higher, but the estimated values for
$\Omega$ and their redshift dependence do not change substantially.

\cite{carlberg96a} studied 16 X-ray selected clusters between $z\sim
0.2$ and $0.5$, in the K-corrected $r$-band. Their estimate of
$\Omega(z\approx 0.3) = 0.29 \pm 0.06$ is not far from what observers
in our simulation would conclude for that redshift, 
$\Omega(z\approx 0.3) \approx 0.38$, using $V$-band
luminosities.  However, the correct value in the simulations is
$\Omega(z=0.3) = 0.56$.
\cite{carlberg96a} found no statistically significant dependence of
either $\Upsilon_r$ or $\Upsilon_{r,c}$ on redshift, but
their errors are large enough to be consistent with the level of
evolution of these quantities seen in the simulations.  We caution,
therefore, that estimates of $\Omega_0$ from cluster mass-to-light
ratios can be affected by bias, by the differential fading between the
cluster galaxies and the field galaxies, and by the evolution of both
these quantities with redshift.


As an aside, Tables \ref{clustersa} and \ref{clustersb} also show that
$\Omega_0$ estimated from the baryonic mass in each cluster, $\Omega_b
M_{\mathrm{tot}}/M_{\mathrm{baryons}}$, is biased slightly high. This effect
occurs because of the slight antibias (5 -- 10\%) of baryons with respect to
mass at low redshift (due, presumably, to shocking and outflows in the gas
which prevent baryons from flowing into the potential wells as efficiently as
dark matter does). This result is consistent with previous work
(\cite{evrard90ap}; \cite{cen93ap}; \cite{white93ap}; \cite{lubin96ap}).
Measuring $M_{\mathrm{baryons}}$ is possible in the clusters because the gas is
hot enough to be visible in X-rays; the main difficulty is in measuring
$\Omega_b$, which is generally based on high-redshift deuterium abundance
measurements (e.g., \cite{burles98ap}) and Big Bang nucleosynthesis
calculations (e.g., \cite{schramm98ap}).
The fact that the baryon mass in clusters is closely related to the total mass
makes this method of estimating $\Omega_0$ less prone to the complications of
hydrodynamics, galaxy formation, and stellar evolution than using mass-to-light
ratios.

\begin{figure}
\figurenum{\fignum}
\epsscale{1.0}
\plotone{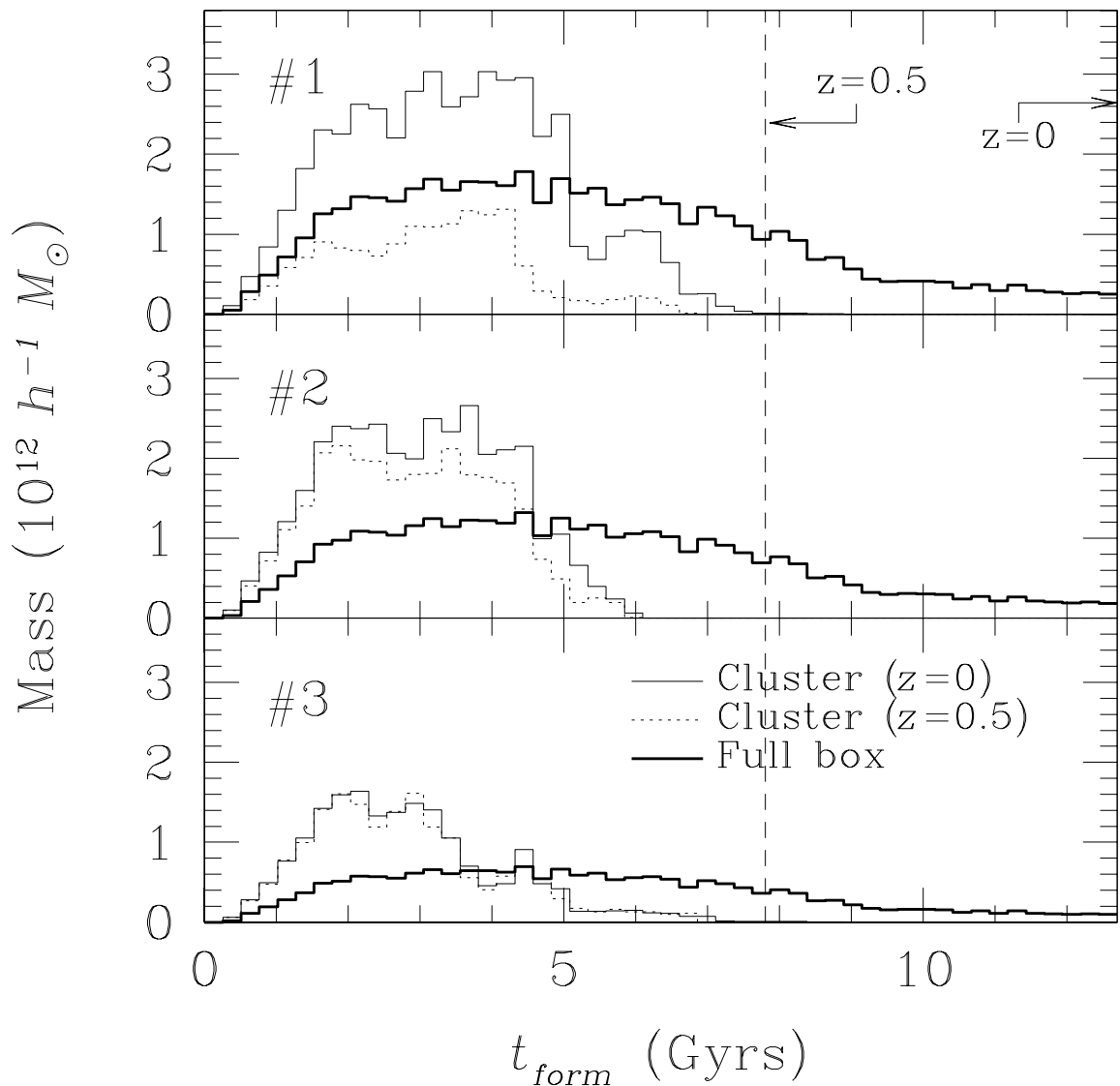}
\epsscale{1.0}
\caption{\label{cgages} Star-formation history in clusters at $z=0$ (thin solid
line) and $z=0.5$ (thin dotted line). Note that most star formation is complete
in these clusters long before the universe is 7.8 Gyrs old, which corresponds
to $z=0.5$ (shown as the vertical dashed line). For this reason, there is no
{\it excess} of blue galaxies at $z=0.5$, although there may be at somewhat
higher redshifts. For comparison, we show the star formation history of the
full box as the thick solid line. This figure assume the instantaneous
burst (IB) approximation.}
\end{figure}

\begin{figure}
\figurenum{\fignum}
\epsscale{1.0}
\plotone{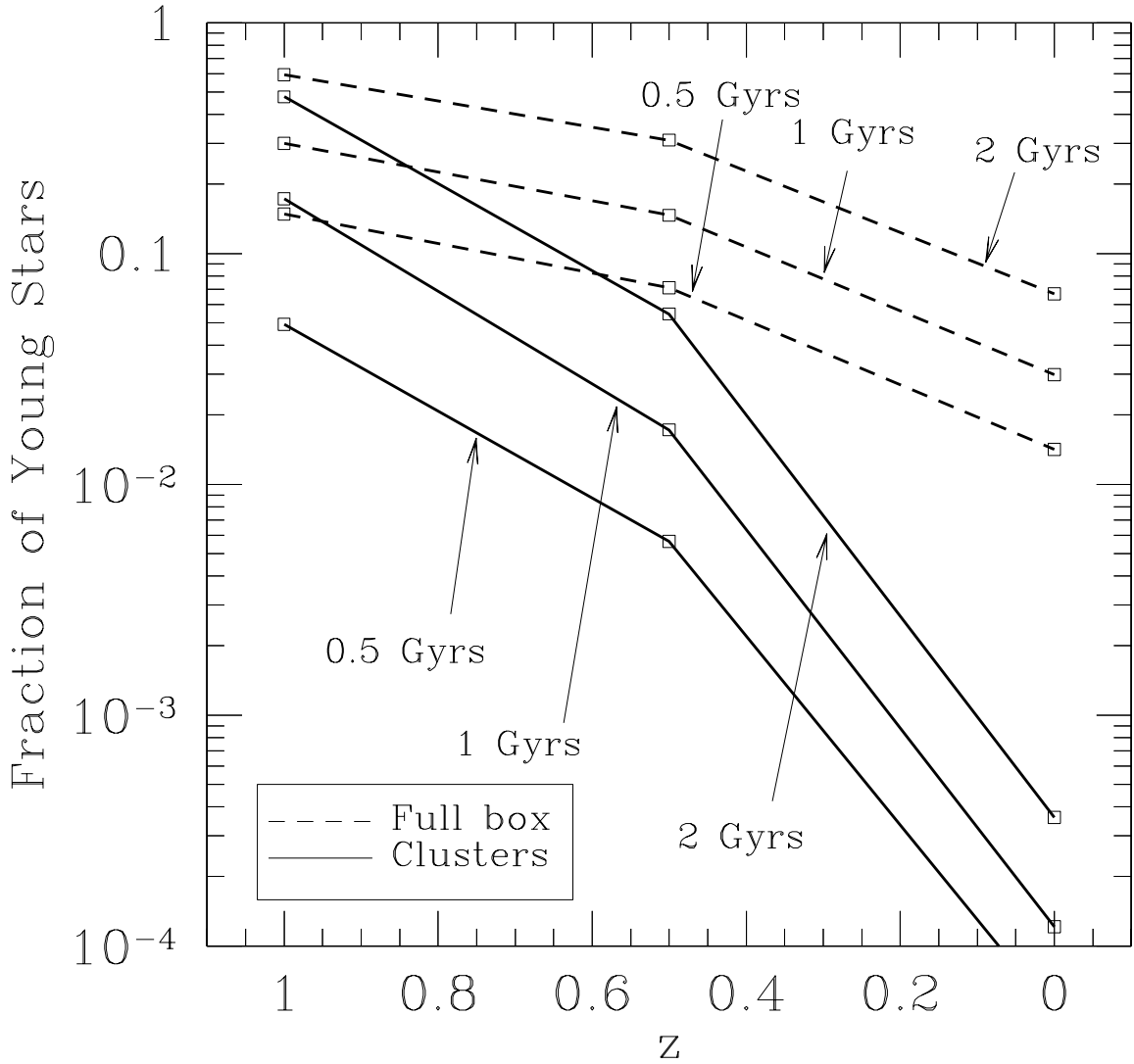}
\epsscale{1.0}
\caption{\label{jerryfig} Mass fraction of recently formed stars in
the clusters (solid lines) and in the whole box (dashed lines), using
the ELS star formation model described in the text. We show results
for stars formed in the previous 0.5, 1, and 2 Gyrs. In each case, the
rapid decline in the clusters compared to the field is evident.}
\end{figure}

From the results of Section \ref{timeformation}, we expect that
cluster galaxies at $z=0.5$ should be younger and bluer than cluster
galaxies at $z=0$. Does this account for the Butcher-Oemler effect ---
the existence of a blue tail in the distribution of $B-V$ colors of
cluster galaxies at $z\sim 0.5$ (Butcher \& Oemler 1978, 1984)?  The
simplest approach to this question is to ask whether the mass ratio of
recently formed galaxies to all galaxies 
in the cluster changes from $z=0$ to
$z=1$. Figure \ref{cgages} shows the distribution of formation times
of galaxy particles in each cluster at $z=0$ and at $z=0.5$, assuming
the IB model of star formation. For comparison, the thick histogram
shows the star formation history of the whole box.  At both $z=0$ and
at $z=0.5$, there has been no star formation in the previous 1 Gyr in
any cluster. On the other hand, the more realistic ELS model allows
star formation to persist somewhat longer.  Figure \ref{jerryfig}
illustrates this effect by showing the fraction of the stars in the
clusters formed in the previous 0.5, 1. and 2 Gyrs at $z=0$, $z=0.5$,
and $z=1$, now using the ELS model. For comparison, we show the same
curves for the whole box. Clearly, the fraction of young stars
increases with redshift more quickly in the clusters than in the
field. This is qualitatively in agreement with the observed
Butcher-Oemler effect.

We can look at the problem in a more observational context by
calculating the $B-V$ colors of cluster galaxies as a function of
redshift.  Following Butcher \& Oemler (1978, 1984), we define the
quantity $f_b$ (the ``blue fraction'' of galaxies) to be the fraction
of mass in galaxy particles whose colors are more than 0.2 mag bluer
than the peak in the $B-V$ distribution.  Observationally, $f_b\sim
0.05$ at $z=0$, but at redshift $z=0.5$, $f_b \sim 0.1$--$0.25$
(Butcher \& Oemler 1978, 1984; \cite{oemler97ap}).  We consider both
the IB and ELS approximations. Under the IB approximation, the colors
redden as the galaxy populations evolve (about 0.1 mag between $z=0.5$
and $z=0$), but the shape of the $B-V$ distribution changes little;
for all the clusters at both $z=0$ and $z=0.5$, $f_b \sim 0.05$ for
the IB approximation.  However, using the ELS model improves the
situation considerably; 
we find $f_b \sim
0.05$ at $z=0$ and $f_b \sim 0.1$--$0.15$ at $z=0.5$, in good agreement
with observations. Thus, these simulations may be revealing the
mechanism behind the Butcher-Oemler effect.

It is worth noting that the astrophysics involved in the star
formation history of cluster galaxies is clearly much more complicated
than that which we model here.  First, the star formation history
inside each galaxy particle we create is surely more complicated than
any the models discussed in the beginning of this section.  Second,
star-burst mechanisms such as ram-pressure induced star formation,
which would affect galaxies falling into clusters
(\cite{dressler90ap}), or merger-induced star formation, which would
affect galaxies in dense environments, could be important. These
mechanisms are not modelled by these simulations.  Third, as stressed
by \cite{kauffmann95a}, the population of clusters observed at $z=0.5$
in an $\Omega = 1$ universe necessarily contains rarer peaks than the
population of clusters observed at $z=0$ (\cite{andreon99ap}), and the
differences in the formation histories of such different objects may
account for the Butcher-Oemler effect. On the other hand, this effect
should be unimportant for a low-density universe, in which structure
forms relatively early.

\setcounter{section}{5}
\section{Conclusions}
\label{conclusions}

We have examined the spatial distribution of galaxy formation as a function of
redshift in a cosmological hydrodynamic simulation of a $\Lambda$CDM universe
with $\Omega_0=0.37$. We quantified the evolution of the clustering of galaxy
formation with the bias $b_\ast(z)$ and the correlation coefficient $r_\ast(z)$
of galaxy formation. In our simulations, $r_\ast(z)$ evolves considerably,
whereas in the peaks-biasing formalism it remains unity at all redshifts.  
The time history of these quantities, combined with the
gravitational debiasing, causes the evolution of $b_g(z)$ and $r_g(z)$ of all
galaxies. As described in Section \ref{observables}, this history of galaxy
formation could be probed observationally by examining the evolution of
$b_{\ast g} r_{\ast g}$, the linear regression of the star formation weighted
galaxy density fields on the number-weighted galaxy density field. The
mass-to-light ratios of clusters are also profoundly affected by this history,
with the effect that while $\Omega(z)$ increases with redshift, the
estimated $\Omega_{\mathrm{est}}(z)$ actually decreases. Finally, the evolution
of the distribution of galaxy formation in the simulation bears a qualitative
relation to the Butcher-Oemler (1978, 1984) effect, and under certain
assumptions about the star formation history of each galaxy particle can even
quantitatively account for it.

Understanding these results completely will require future work.
First, the low resolution of these hydrodynamic simulations (relative to
state-of-the-art $N$-body calculations) calls into question the
accuracy with which we can follow the process of galaxy formation, and
makes us unable to identify individual galaxies in high-density
regions. Thus we cannot follow adequately the effects of the merging
and destruction of halos, especially in clusters. This fact was the
basis of our decision to examine the stellar {\it mass} density field
rather than the galaxy {\it number} density field. Observationally, it
is possible to trace the stellar mass density crudely by weighting the
galaxy density field by luminosity, especially at longer
wavelengths. The evolution of bias in the 
luminosity-weighted galaxy density field is dominated by the three
mechanisms described in Section \ref{motiv}. The evolution of bias
in the {\em number}-weighted galaxy density field is additionally affected
by the process of merging in dense regions.  We do not have the
resolution to treat this effect here; \cite{kravtsov98a} explore this
using a high-resolution $N$-body code. 

Equally important is to explore the dependence of the evolution of
$b_\ast(z)$ and $r_\ast(z)$ on cosmology.  As we saw in Section
\ref{timeformation}, their evolution was dominated by the evolution of
the density-temperature relationship; as this relationship and its
time-dependence varies from cosmology to cosmology, we expect
$b_\ast(z)$ and $r_\ast(z)$ to vary as well. Just as important is the
uncertainty in the star formation model used, whose properties are
evidently rather important in determining the clustering and
correlation properties of the galaxies. 
Furthermore, understanding the statistical properties of galaxy
clusters in the simulations will require more realizations than the
single one we present. Finally, larger volumes than we simulate will
be necessary to compare our results to the large redshift surveys,
such as the SDSS, currently in progress.

Since these simulations are expensive and time-consuming to run, we would like
to find less expensive ways of increasing the volume of parameter space and
real space which we investigate. It may be possible to capture the important
aspects of the hydrodynamic code by using much less expensive $N$-body
simulations. One way to do so is to determine the relationship between galaxy
density, mass density, and velocity dispersion at $z=0$, and simply apply that
relation to the output of an $N$-body code at redshift zero, as suggested by
\cite{blanton98a}. However, 
the
final relationship between galaxies, mass, and temperature will depend both
on the time evolution of the mass-temperature relationship, and the
gravitational debiasing. For instance, if all galaxies form at high
redshift, galaxies will have had 
time to
debias completely. To take into account such effects explicitly, we want to
apply a galaxy formation criterion at each time step, and follow the
gravitational evolution of the resulting galaxies to find the final
distribution at $z=0$.  If one could do so, one could run large volume dark
matter simulations, in which one formed galaxies in much the same way as they
form here. Motivated by Figure \ref{dTg}, we propose applying our measured
$\avg{\delta_\ast|\delta, T}$ on 1 $h^{-1}$ Mpc scales in order to form
galaxies in such an $N$-body simulation, using the local dark-matter velocity
dispersion or the local potential as a proxy for temperature. This approach
assumes only that the dependence of galaxy formation on local density and local
temperature is not a strong function of cosmology. This method would be similar
in spirit to, but simpler than, the semi-analytic models which have been
implemented in the past.

The results of Section \ref{burst} suggest a simple way of exploring the
effects of changing the galaxy formation criteria in such $N$-body simulations,
or for that matter in the hydrodynamical simulations themselves, as long as one
is content with studying second moments and with ignoring the effects of the
feedback of star formation. Given more output times than we have available from
these simulations, one can simply determine the location of galaxy formation at
each time step -- {\it after the fact} -- and use Equation (\ref{fullev}) to
propagate the results to $z=0$, or to whichever redshift one wants. Thus, one
could easily examine the effects that varying the galaxy formation criteria
would have. Compared to rerunning the full simulation, it is computationally
cheap to determine the properties of $\delta_\ast$ at each redshift (given the
galaxy formation model) and then to reconstruct the evolution of $b_g(z)$ and
$r_g(z)$. 

The continuity equation may also prove useful in comparing observations to
models. For instance, large, deep, angular surveys in multiple bands, such as
the SDSS, will permit the cross-correlation of different galaxy types as a
function of redshift, using photometry to estimate both redshift and galaxy
type (\cite{connolly98ap}). If one properly accounts for the evolution of
stellar colors, one will be able to test the hypothesis that two galaxy types
both obey the continuity equation. The sense and degree of any discrepancy will
shed light on whether one type of galaxy merges to form another, or whether new
galaxies of a specific type are forming.

Estimates of the star formation rate as a function of redshift are now possible
(\cite{madau96ap}). Perhaps the next interesting observational goal is to study
the evolution of the clustering of star formation. The Butcher-Oemler effect
tells us something about the evolution of star formation in dense regions, but
it is necessary to understand the corresponding evolution in the field, as
well.  The correlation of star-forming galaxies at different redshifts with
the full galaxy population may be a more sensitive discriminant
between galaxy formation models than measures of the clustering
of all galaxies. 

\acknowledgments

This work was supported in part by the grants NAG5-2759, NAG5-6034,
AST93-18185, AST96-16901, and the Princeton University Research
Board. MAS acknowledges the additional support of Research
Corporation. MT is supported by the Hubble Fellowship HF-01084.01-96A
from STScI, operated by AURA, Inc., under NASA contract NAS5-26555.
We would like to thank James E.~Gunn, Chris McKee, Kentaro Nagamine,
and David N.~Spergel for useful discussions, as well as St\'ephane
Charlot for kindly providing his spectral synthesis results.

\appendix

%
%
%
\section{Understanding Galaxy Formation in the Simulation}
\label{formation}

The conditions required for galaxy formation in each cell of the
simulation are:
\begin{eqnarray}
\delta &>& 5.5,\cr
m_{\mathrm{b}} &>& m_{\mathrm{J}} \equiv G^{-3/2} \rho_{\mathrm{b}}^{-1/2}
C^3 \left[1 +\frac{\delta_d+1}{\delta_b+1} 
\frac{\Omega_d}{\Omega_b}\right]^{-3/2},\cr
t_{\mathrm{cool}} &<& t_{\mathrm{dyn}} \equiv \sqrt{\frac{3\pi}{32
G\rho}} \mathrm{,~and} \cr
\nabla\cdot\vv{v} &<& 0.
\end{eqnarray}
The subscript $b$ refers to the baryonic matter; the subscript $d$
refers to the collisionless dark matter.  In words, these criteria say
that the overdensity must be reasonably high, the mass in the cell
must be greater than the Jeans mass, the gas must be cooling faster
than the local dynamical time, and the flow must be converging.
Assuming that the time scale for collapse is the dynamical
time, we transfer mass from the gas to collisionless particles at the rate:
\begin{equation} 
\label{weighting}
\frac{\Delta m_\ast}{\Delta t} = \frac{m_b}{t_\ast}.
\end{equation} 
where $t_\ast=t_{\mathrm{dyn}}$ if $t_{\mathrm{dyn}} > 100$ Myrs and
$t_\ast= 100$ Myrs if $t_{\mathrm{dyn}} > 100$ Myrs.

\begin{figure}
\figurenum{\fignum}
\epsscale{1.0}
\plotone{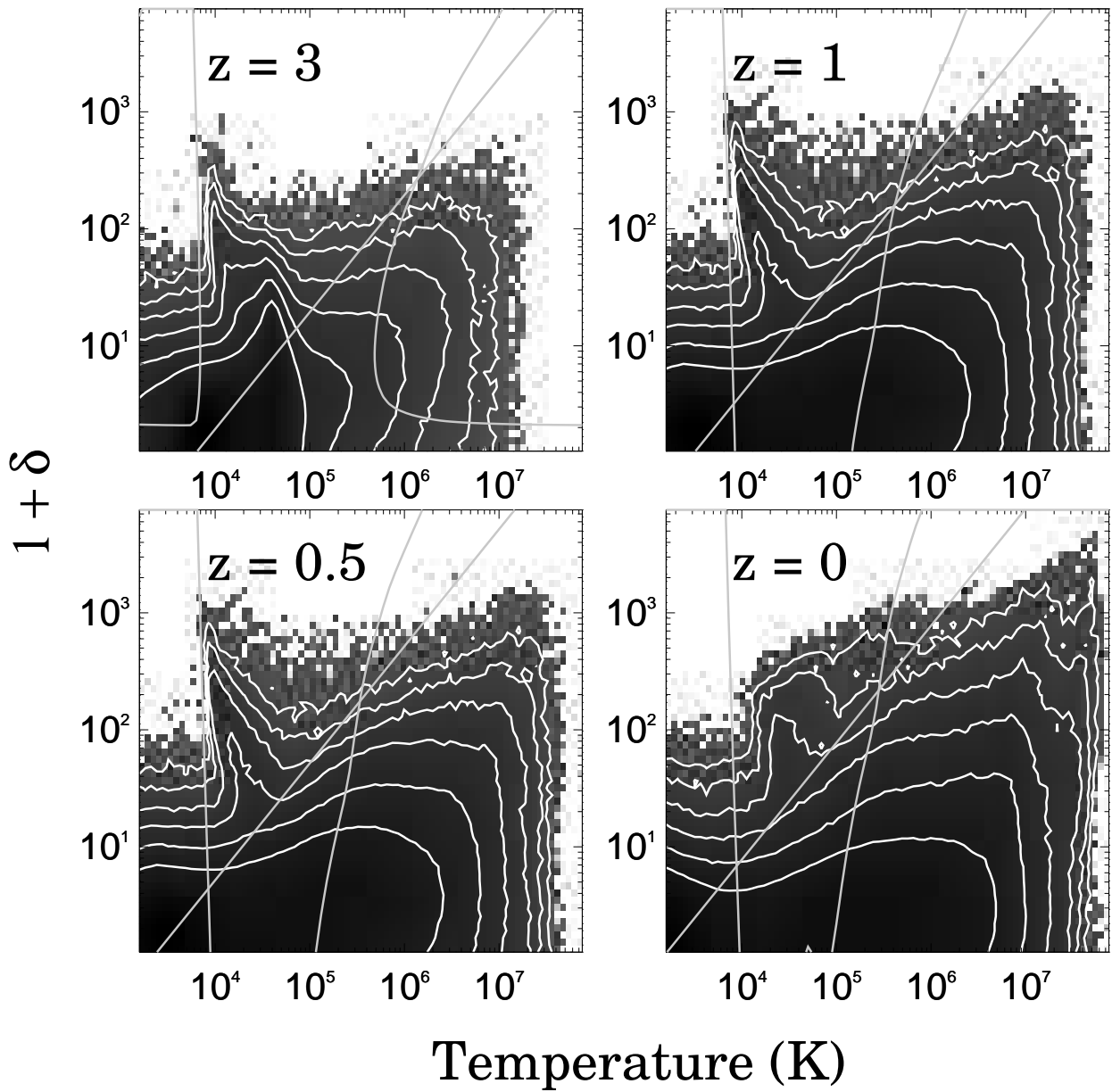}
\epsscale{1.0}
\caption{\label{smalldT} Volume-weighted joint distribution of mass
density and temperature at 0.2 $h^{-1}$ Mpc scales (the grid cell size
of the simulation), at each output redshift.  Greyscale and contours
are as in Figure \ref{dT}. Note that at these scales, the peak in the
cooling curve at $10^4$ to $10^5$ K is noticeable as a ridge at around
$1+\delta\sim 100$. We overlay the Jeans criterion for galaxy
formation as the diagonal line and the cooling criterion (for
one-percent solar metallicity) as the two vertical curves. The
simulation produces galaxies above the Jeans line and in between the
two cooling lines. At all redshifts, the densest regions 
automatically satisfy the cooling criterion as long as they satisfy
the Jeans criterion.}
\end{figure}

To understand the effects of these criteria, consider Figure
\ref{smalldT}, which shows the distribution of $1+\delta$ and $T$ at
the scale of a single cell at each redshift.  We show the Jeans mass
criterion as the diagonal grey line (points to the left are Jeans
unstable), and the cooling criterion (for one-percent solar
metallicity gas) as the curved grey lines (points in between the lines
can cool efficiently). For these
purposes we have assumed $\delta_b=\delta$, $\gamma=5/3$, and
$\mu=m_H$. 
Note that at high densities, satisfying
the Jeans criterion automatically satisfies the cooling
criterion. At high redshifts, there is a fair amount of gas which
appears to have cooled
into this regime, at $10^4$--$10^5$ K and at relatively high
densities. As the density fields evolve,
the densest regions become hotter, making it more difficult for the gas
there to cool enough to allow galaxy formation.

\clearpage
%
%

\setcounter{thefigs}{0}

\clearpage
\stepcounter{thefigs}


\clearpage
\stepcounter{thefigs}

\clearpage
\stepcounter{thefigs}

\clearpage
\stepcounter{thefigs}


\clearpage
\stepcounter{thefigs}

\clearpage
\stepcounter{thefigs}

\clearpage
\stepcounter{thefigs}

\clearpage
\stepcounter{thefigs}

\clearpage
\stepcounter{thefigs}

\clearpage
\stepcounter{thefigs}

\clearpage
\stepcounter{thefigs}

\clearpage
\stepcounter{thefigs}

\clearpage
\stepcounter{thefigs}


\clearpage
%
%

\setcounter{thetabs}{0}

\stepcounter{thetabs}
\begin{deluxetable}{ccccccc}
\tablecolumns{7}
\tablenum{\tabnum}
\tablecaption{\label{clustersa} Clusters at $z=0$, at which time
$\Omega=0.37$.}
\tablehead{ & $M_{\mathrm tot}$ ($h^{-1}$ $M_\odot$) 
& $\avg{v^2}^{1/2}$ (km/s) & 
$\Upsilon_B/\Upsilon_{B,c}$ & 
$\Upsilon_V/\Upsilon_{V,c}$ & $\Upsilon_K/\Upsilon_{K,c}$ & 
$\Omega_{b0} M_{\mathrm{tot}}/M_{\mathrm{baryons}}$}
\startdata
\#1 & $9.847\times 10^{14}$ 
& 1076.505 & 
0.414 & 0.366 & 0.316 & 0.390 \cr
\#2 & $7.919\times 10^{14}$ 
& 951.378 & 
0.442 & 0.401 & 0.359 & 0.401 \cr
\#3 & $5.448\times 10^{14}$ 
& 832.657 & 
0.535 & 0.498 & 0.481 & 0.409 \cr
\enddata
\end{deluxetable}

\stepcounter{thetabs}
\begin{deluxetable}{ccccccc}
\tablecolumns{7}
\tablenum{\tabnum}
\tablecaption{\label{clustersb} Clusters at $z=0.5$, at which time
$\Omega=0.66$.} 
\tablehead{ & $M_{\mathrm tot}$ ($h^{-1}$ $M_\odot$) 
& $\avg{v^2}^{1/2}$ (km/s) & 
$\Upsilon_B/\Upsilon_{B,c}$ & 
$\Upsilon_V/\Upsilon_{V,c}$ & $\Upsilon_K/\Upsilon_{K,c}$ & 
$\Omega_{b0} M_{\mathrm{tot}}/M_{\mathrm{baryons}}$}
\startdata
\#1 & $2.698 \times 10^{14}$ 
& 751.243 & 
0.321 & 0.275 & 0.231 & 0.716 \cr
\#2 & $6.468 \times 10^{14}$ 
& 1135.564 &
0.451 & 0.390 & 0.321 & 0.719 \cr
\#3 & $3.839 \times 10^{14}$ 
& 815.936 & 
0.397 & 0.349 & 0.304 & 0.663 \cr
\enddata
\end{deluxetable}

\end{document}